\documentclass[
superscriptaddress,
preprint,
 amsmath,amssymb,
 aps,
]{revtex4-2}

\usepackage{float} 
\usepackage{graphicx}
\usepackage{dcolumn}
\usepackage{bm}
\usepackage{nicefrac}
\usepackage{booktabs} 
\usepackage{multirow}
\usepackage{bigdelim}
\usepackage{listings}
\usepackage{tabularx}
\usepackage{physics}
\usepackage{amsmath}
\usepackage{dsfont}
\usepackage{amssymb}
\usepackage{bbold}
\usepackage{subfigure}
\usepackage{microtype}
\usepackage[version=4,arrows=pgf]{mhchem}
\usepackage{booktabs}
\usepackage{xcolor}
\usepackage{siunitx} 
\usepackage{mathtools}
\usepackage{lipsum}
\usepackage[T1]{fontenc}

\begin{document}


\title{Two-Photon Interference of Single Photons from Dissimilar Sources}

\author{Christian Dangel}
\thanks{These two authors contributed equally to this work.}
\author{Jonas Schmitt}
\thanks{These two authors contributed equally to this work.}
\affiliation{Walter Schottky Institut and Physik Department, Technische Universit\"at M\"unchen, Am Coulombwall 4, 85748 Garching, Germany}
\affiliation{Munich Center for Quantum Science and Technology (MCQST), Schellingstr. 4, 80799 Munich, Germany}
\author{Anthony J. Bennett}
\affiliation{School of Engineering, Cardiff University, Cardiff, CF24 3AA, United Kingdom}
\author{Kai M\"uller}
\affiliation{Munich Center for Quantum Science and Technology (MCQST), Schellingstr. 4, 80799 Munich, Germany}
\affiliation{Walter Schottky Institut and Department of Electrical and Computer Engineering, Technische Universit\"at M\"unchen, Am Coulombwall 4, 85748 Garching, Germany}
\author{Jonathan J. Finley}
\email{finley@wsi.tum.de}
\affiliation{Walter Schottky Institut and Physik Department, Technische Universit\"at M\"unchen, Am Coulombwall 4, 85748 Garching, Germany}
\affiliation{Munich Center for Quantum Science and Technology (MCQST), Schellingstr. 4, 80799 Munich, Germany}

\date{\today}

\begin{abstract}
Entanglement swapping and heralding are at the heart of many protocols for distributed quantum information. For photons, this typically involves Bell state measurements based on two-photon interference effects. In this context, hybrid systems that combine high rate, ultra-stable and pure quantum sources with long-lived quantum memories are particularly interesting. Here, we develop a theoretical description of pulsed two-photon interference of photons from dissimilar sources to predict the outcomes of second-order cross-correlation measurements. These are directly related to, and hence used to quantify, photon indistinguishability. We study their dependence on critical system parameters such as quantum state lifetime and frequency detuning, and quantify the impact of emission time jitter, pure dephasing and spectral wandering. Our results show that for fixed lifetime of emitter one, for each frequency detuning there is an optimal lifetime of emitter two that leads to highest photon indistinguishability. Expectations for different hybrid combinations involving III-V quantum dots, color centers in diamond, 2D materials and atoms are quantitatively compared for real-world system parameters. Our work both provides a theoretical basis for the treatment of dissimilar emitters and enables assessment of which imperfections can be tolerated in hybrid photonic quantum networks.

\end{abstract}

\maketitle


\section{\label{sec:intro} Introduction}



\noindent Two-photon interference lies at the heart of entanglement swapping and is a central component needed for distributed quantum technologies\cite{insideqreps, barrettkok, repeatuntilsuccess, network_atom2}. 
In the context of quantum communication, key steps have recently been made toward establishing real world quantum links using photons to entangle one or more atoms in high performance cavity QED systems \cite{langenfeld2021quantum}. However, the principle challenges that must be overcome to extend the length of quantum channels are absorption in optical fibers, and decoherence in the static quantum memories that store quantum information during classical communication, measurement processing and error correction \cite{vanlook}.
Long-range networks have been demonstrated using trusted nodes \cite{trustednodes1}, but different approaches are needed for unconditionally secure links, necessitating the development of quantum repeaters \cite{vanlook, qrepeaters}. 
The simplest repeater schemes involve two quantum sources located at nodes A and B, each emitting single photons that are entangled with one of their internal degrees of freedom \cite{vanlook}. These photons are typically sent to an intermediate central node where a Bell state measurement is performed to swap entanglement between the communicating parties\cite{bsm,her_ent_atoms_tpi}. The use of quantum memories at the intermediate node allows for variable photon arrival times\cite{naturelukin} and, moreover, it permits measurement-dependent local unitary operations and quantum error correction protocols \cite{purification,qec} to be performed. 

%
Key factors that impact upon the efficiencies of such quantum links are deterministic single photon sources operating at high rates, as well as memories with near unity photon in- and out-coupling efficiencies and long coherence times \cite{vanlook}.  Amongst all the quantum systems studied to date, trapped atoms \cite{network_atom1, network_atom2, langenfeld2021quantum} or ions \cite{network_ion, network_ions2} have probably made the most impressive demonstrations. However, solid-state approaches may provide routes towards integration and scalability.  In this context, paramagnetic defects in diamond \cite{network_defect1, network_defect2} or 2D materials \cite{network_2d0, network_2d}, semiconductor quantum dots (QDs) \cite{network_qd, network_qd2}, rare-earth ions\cite{gritsch2021narrow} and superconducting qubits \cite{network_sc1, network_sc2} each have specific advantages and disadvantages. Unfortunately, none of these systems presents both ideal source and memory characteristics and hybrid schemes that combinine the beneficial properties of different platforms are becoming increasingly interesting \cite{vanlook, hybrid}. Of all the potential systems, QDs have proven to be the best emitters, as manifested by their high brightness, large clock rates approaching 1 GHz \cite{vanlook}, excellent single photon purity ($g^2(0)\leq10^{-3}$) and quantum indistinguishability \cite{senderqd0}. Very recently, end to end system efficiencies up to 57\% \cite{57eff} have been demonstrated using InAs QDs in point-to-point links. Other important features are emission frequency control\cite{qdsreview} that allows matching to other types of emitter and their ability to emit quantum light in the telecommunications C and O-bands. However, the Achilles heel of QDs is the comparatively short coherence times of electron and hole spin qubits (typically $\leq 1~\mu$s \cite{warburton2013single}). In comparison, spin coherence times for negatively charged silicon vacancy centers in diamond can be $\geq10$ ms \cite{siv_13mscoh} and recent advances in the processing of diamond nano-photonic structures has also led to very impressive (near deterministic) spin-photon interfaces \cite{memorysiv}. First proof-of-principle experiments have already demonstrated the functionality of diamond in repeater architectures \cite{memorysiv0, memorysiv2} and protocols exist for photon based quantum information processing using diamond \cite{nemoto2014photonic}.\\ 

In order to perform high-fidelity entanglement swapping between photons emitted by two different quantum systems, their wavepackets 
must be indistinguishable in their first order coherence properties as well as in their spatio-temporal profile, including polarization \cite{simulationpaper}.
Thus, in this paper we develop a theoretical model to describe two-photon interference from dissimilar sources and use our model to predict the results of experiments performed on hybrid quantum systems. The sources will be characterized by system properties such as excitation pulse width and temporal form, spontaneous emission decay lifetime and jitter of the emission frequency arising from cascaded emission processes and spectral wandering. By simulating the outcomes of two-photon interference experiments in a general way, we obtain results that are applicable to different combinations of quantum emitters. As such, our results thus serve as a basis to guide different hybrid quantum repeater implementations where Bell state measurements are performed on photon pairs generated at different locations.

\section{Quantification of Photon Indistinguishability}

\begin{figure}[H]
	\centering
	\includegraphics[width=0.5\linewidth]{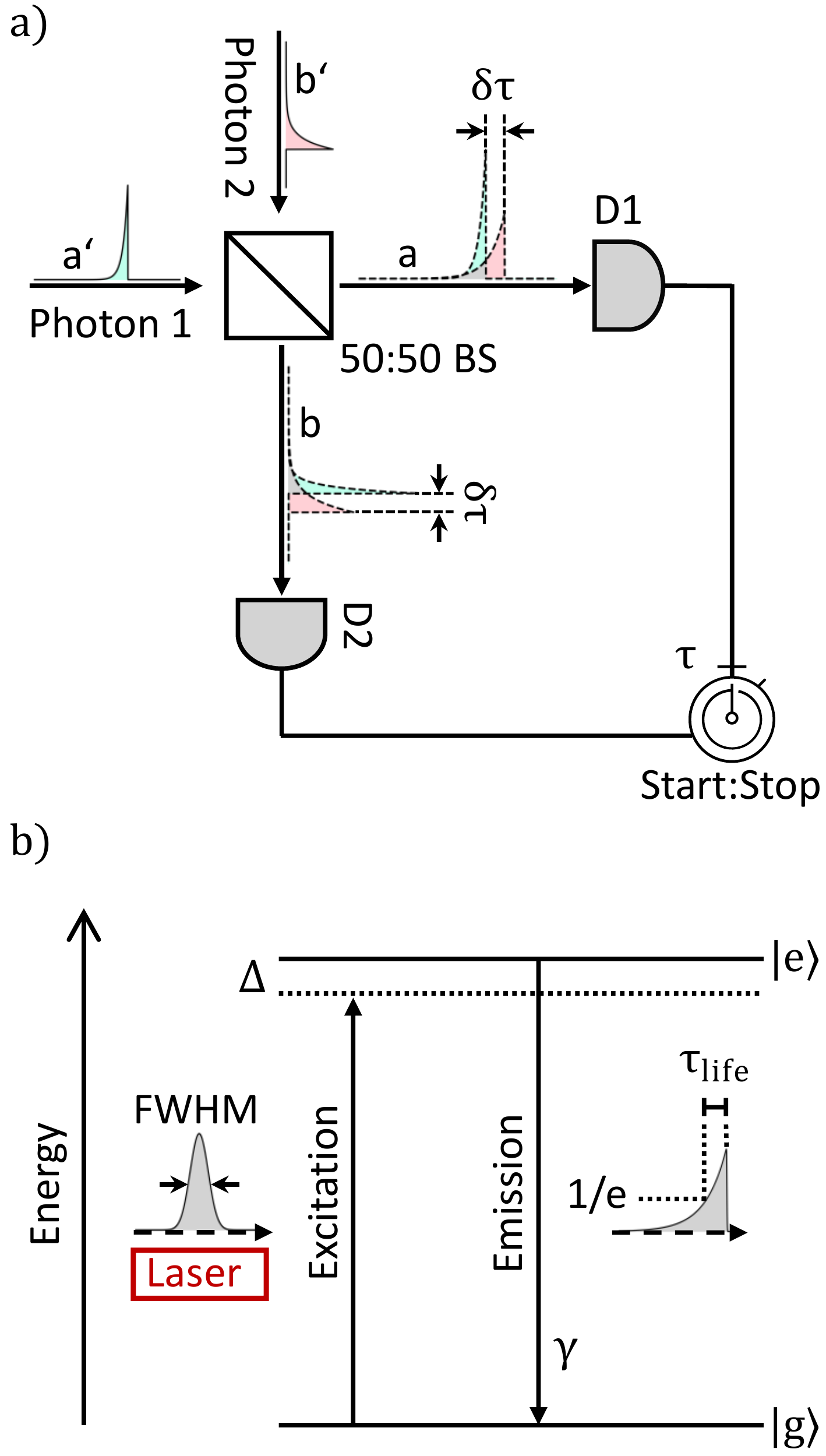}
	\caption{(a) Schematic setup of a typical HOM type experiment for the investigation of photon indistinguishability with $a'$ and $b'$ being the input light fields and a and b the light fields after the beamsplitter. A potential relative delay in photon arrival time is denoted by $\delta \tau$, while $\tau$ represents the interval between coincident signals at the different detectors D1 and D2. (b) Excitation of a two-level system. A Gaussian $\pi$-pulse of width $\tau_{pulse}$ brings the two-level system from its ground state $\ket{g}$ to its excited state $\ket{e}$. The laser frequency may be detuned by an amount $\Delta$. The spatio-temporal profile of the emitted photon is governed by the emitter decay rate $\gamma = 1/\tau_{life}$.} \label{fig:setup}
\end{figure}

Indistinguishability of single photons is most commonly investigated via Hong-Ou-Mandel (HOM) \cite{hom_original, hom2orig} type two-photon interference experiments. 
Figure \ref{fig:setup}a illustrates the typical measurement scenario: Two photons propagating in modes $a'$ and $b'$, at the input of the 50:50 beamsplitter with a potential relative temporal delay $\delta \tau$. The transformed light fields $a$ and $b$ are then recorded by photon counters that correlate coincidence counts as a function of the time delay $\tau$ elapsing between a start signal at detector D1 and a subsequent stop signal at D2. Upon forming a temporal average over many input photon pairs, a histogram is obtained that reflects the number of coincident detections for each time interval $\tau$. Coincidences near $\tau=0$ correspond to simultaneously arriving photons at both detectors and thus label distinguishable photons. In contrast, the absence of coincidences around $\tau=0$ is a signature of HOM coalescence and thus of photon indistinguishability.  

In this paper we develop a theoretical model that provides access to pre-defined quantum mechanical operations as well as representations of states and operators in matrix form and implement it using the quantum toolbox in Python \cite{qutip}. In particular, we evaluate two-time correlators of the form $\expval*{A(t)B(t+\tau)C(t)}$ using built-in functions based on an extended form of the quantum regression theorem \cite{qregthm} as implemented by Kevin Fischer et al. \cite{simulationpaper}.
\noindent We begin by developing a general expression for the intensity cross-correlation function measured in a HOM experiment $G^{(2)}_{HOM}(t, \tau)$. This quantifies correlations of the fields at the two detectors and corresponds to the joint probability density of detecting a photon at detector one at time $t$ and detecting a second photon at detector two at time $t+\tau$. In the case of perfectly indistinguishable single photons, $G^{(2)}_{HOM}(t, \tau)$ is zero for any $t$ and $\tau$, since the photons always exit the beamsplitter together at either of its output ports. In the most general case, the non-normalized intensity cross-correlation function of two quantized fields is given by \cite{g1g2_scully} 
\begin{eqnarray}
G^{(2)}_{ab}(t, \tau) = \expval*{\hat{b}^{\dagger}(t)\hat{a}^{\dagger}(t + \tau)\hat{a}(t +\tau) \hat{b}(t)} . \label{eq:g2HOM_definition}
\end{eqnarray}
\noindent As depicted schematically in fig. \ref{fig:setup}a we note that $\hat{a}$ and $\hat{b}$ in eqn. \ref{eq:g2HOM_definition} are the fields at the detectors at the output ports of the beamsplitter. To establish a connection to the underlying system dynamics, we express them in terms of the input fields $\hat{a}'$ and $\hat{b}'$ via the usual beampsplitter unitary transformation \cite{haroche_raimond}.
Substituting them, eqn. \ref{eq:g2HOM_definition} can be written in terms of the creation and annihilation operators of the input field modes:
\begin{widetext}
\begin{equation}
\begin{aligned}
&~~~G^{(2)}_{HOM}(t, \tau) = \frac{1}{4}  \cdot \\
&~~(\expval*{\hat{a}'^{\dagger}(t)\hat{a}'^{\dagger}(t+\tau)\hat{a}'(t+\tau)\hat{a}'(t)}  
+\expval*{\hat{b}'^{\dagger}(t)\hat{b}'^{\dagger}(t+\tau)\hat{b}'(t+\tau)\hat{b}'(t)} 
\rightarrow \text{(i)}\\
&+ \expval*{\hat{a}'^{\dagger}(t)\hat{b}'^{\dagger}(t+\tau)\hat{b}'(t+\tau)\hat{a}'(t)}
+ \expval*{\hat{b}'^{\dagger}(t)\hat{a}'^{\dagger}(t+\tau)\hat{a}'(t+\tau)\hat{b}'(t)} \rightarrow \text{(ii)}\\
&- \expval*{\hat{a}'^{\dagger}(t)\hat{b}'^{\dagger}(t+\tau)\hat{a}'(t+\tau)\hat{b}'(t)} 
-\expval*{\hat{b}'^{\dagger}(t)\hat{a}'^{\dagger}(t+\tau)\hat{b}'(t+\tau)\hat{a}'(t)} 
\rightarrow \text{(iii)}\\
&\begin{rcases}
+ \expval*{\hat{a}'^{\dagger}(t)\hat{a}'^{\dagger}(t+\tau)\hat{a}'(t+\tau)\hat{b}'(t)} 
- \expval*{\hat{a}'^{\dagger}(t)\hat{a}'^{\dagger}(t+\tau)\hat{b}'(t+\tau)\hat{a}'(t)}\\
- \expval*{\hat{a}'^{\dagger}(t)\hat{a}'^{\dagger}(t+\tau)\hat{b}'(t+\tau)\hat{b}'(t)} 
- \expval*{\hat{a}'^{\dagger}(t)\hat{b}'^{\dagger}(t+\tau)\hat{a}'(t+\tau)\hat{a}'(t)}\\
+\expval*{\hat{a}'^{\dagger}(t)\hat{b}'^{\dagger}(t+\tau)\hat{b}'(t+\tau)\hat{b}'(t)} 
+ \expval*{\hat{b}'^{\dagger}(t)\hat{a}'^{\dagger}(t+\tau)\hat{a}'(t+\tau)\hat{a}'(t)}\\
-\expval*{\hat{b}'^{\dagger}(t)\hat{a}'^{\dagger}(t+\tau)\hat{b}'(t+\tau)\hat{b}'(t)}
-\expval*{\hat{b}'^{\dagger}(t)\hat{b}'^{\dagger}(t+\tau)\hat{a}'(t+\tau)\hat{a}'(t)}\\
 -\expval*{\hat{b}'^{\dagger}(t)\hat{b}'^{\dagger}(t+\tau)\hat{a}'(t+\tau)\hat{b}'(t)} 
+\expval*{\hat{b}'^{\dagger}(t)\hat{b}'^{\dagger}(t+\tau)\hat{b}'(t+\tau)\hat{a}'(t)}
) .\end{rcases}\text{(iv)} 
 \label{eq:g2HOM_long}
\end{aligned}
\end{equation}
\end{widetext}

\noindent By considering the two input fields to be independent (i.e. not entangled), $\expval*{\hat{a}'\hat{b}'} = \expval*{\hat{a}'} \expval*{\hat{b}'}$, and noting that $\comm*{\hat{a}'}{\hat{b}'} = \comm*{\hat{a}'^{\dagger}}{\hat{b}'^{\dagger}} = 0$, the expression in eqn. \ref{eq:g2HOM_long} can be grouped into four different types of terms (i)-(iv) \cite{citationkai}. Terms (i), (ii) and (iii) correspond to intensity auto-correlations, intensity two-time correlators and products of field correlators, respectively. Unlike terms (i)-(iii), terms of type (iv) contain a different number of creation and annihilation operators for each field, such that phase factors do not cancel. Considering a realistic scenario where averages are taken over multiple repetitions of an experiment, the random phases cause these terms to average to zero in the temporal average \cite{citationkai}.
The other terms are non-zero for general input states. Discarding the phase-dependent terms (iv) and using the fact that $G^{(1)}_{x'x'}(t,\tau)^* = \expval*{\hat{x'}^{\dagger}(t + \tau) \hat{x'}(t) }$ and $\Re{z} = \frac{1}{2} (z+z^*)$ for any $z \in \mathbb{C}$, eqn. \ref{eq:g2HOM_long} simplifies to the degree of \textit{HOM coherence}:

\begin{eqnarray}
G^{(2)}_{HOM}(t, \tau) = &&  \frac{1}{4} \cdot (G^{(2)}_{a'a'}(t, \tau) + G^{(2)}_{b'b'}(t, \tau) \nonumber \\
&&+ \expval*{\hat{n}_{a'}(t)}\expval*{\hat{n}_{b'}(t+\tau)} +\expval*{\hat{n}_{b'}(t)}\expval*{\hat{n}_{a'}(t+\tau)} \nonumber \\
&&- 2 \Re{G^{(1)}_{a'a'}(t, \tau) G^{(1)}_{b'b'}(t,\tau)^*}) ,\label{eq:ghomttau}
\end{eqnarray}

\noindent where $G^{(1)}_{x'x'}(t, \tau)$ and $G^{(2)}_{x'x'}(t, \tau)$ are the first and second order auto-correlation functions and $\hat{n}_{x'}$ is the number operator of the respective input field $x' \in \{a',b'\}$. The correlators are given by

\begin{eqnarray}
G^{(1)}_{x'x'}(t,\tau) &=& \expval*{\hat{x}'^{\dagger}(t)\hat{x}'(t+\tau)}, \label{eq:ghomttauterms1}
\end{eqnarray} 
where
\begin{eqnarray}
\expval*{\hat{n}_{x'}(t)} &=& \expval*{\hat{x}'^{\dagger}(t)\hat{x}'(t)} \label{eq:ghomttauterms2} 
\end{eqnarray}
and 
\begin{eqnarray}
G^{(2)}_{x'x'}(t,\tau) &=& \expval*{\hat{x}'^{\dagger}(t)\hat{x}'^{\dagger}(t+\tau)\hat{x}'(t+\tau)\hat{x}'(t)}.
\label{eq:ghomttauterms3}
\end{eqnarray}


\noindent Eqns. \ref{eq:ghomttau}-\ref{eq:ghomttauterms2} provide the general framework we use to calculate the results of two-photon interference experiments, given that one measures the first and second order correlators, as well as the intensities of the incident light fields in experiments. 
While the equations are generally valid for both pulsed and continuous wave excitation, we focus on the former case since it is most relevant for a description of deterministic single photon generation.

Figure \ref{fig:setup}b schematically illustrates the scenario described by our simulations. A Gaussian laser pulse of width $\tau_{pulse}$ couples the ground state $\ket{g}$ and excited state $\ket{e}$ of a two-level quantum system (TLS). We focus on two-level systems since they are representative of generic quantum emitters used in quantum networks, representing two specific quantum states in a more general ladder decay scenario. However, the formalism presented in this paper also allows the implementation of more complex systems with multiple eigenstates by using the appropriate Hamiltonian. Generally, we consider resonant excitation but allow for laser detuning such that the excitation frequency can be detuned by $\Delta$. When the system is in its excited state, it emits a photon by decaying to the ground state at a rate $\gamma=1/\tau_{life}$. For Gaussian excitation pulses of width FWHM $ \ll \tau_{life}$ (for details, see Appendix \ref{sec:gauss_ex}), the probability density for emitting a photon decays exponentially with time, which translates into an exponential photon wavepacket in the spatiotemporal domain.

\noindent We continue to explore the time-dependence of the degree of HOM coherence. So far, the expression $G^{(2)}_{HOM}(t,\tau)$ depends on the two different times $t$ and $\tau$. However, in experiments, one is typically not interested in the time $t$ at which the first timer is started, but rather in a histogram for detection time differences $\tau$, where each time bin implicitly comprises all possible values of $t$ for the first detection. We obtain the corresponding probability density function, which we call the \textit{time-resolved degree of HOM coherence}, by integrating $G^{(2)}_{HOM}(t,\tau)$, as defined in equation \ref{eq:ghomttau}, over all possible values of $t$ \cite{citationkai}:

\begin{eqnarray}
G^{(2)}_{HOM}(\tau) &\equiv& \int_{0}^{\infty} dt ~ G^{(2)}_{HOM}(t, \tau) . \label{eq:g2HOM_timeresolved}
\end{eqnarray}	

\begin{figure}
	\centering
	\includegraphics[width=0.6\linewidth]{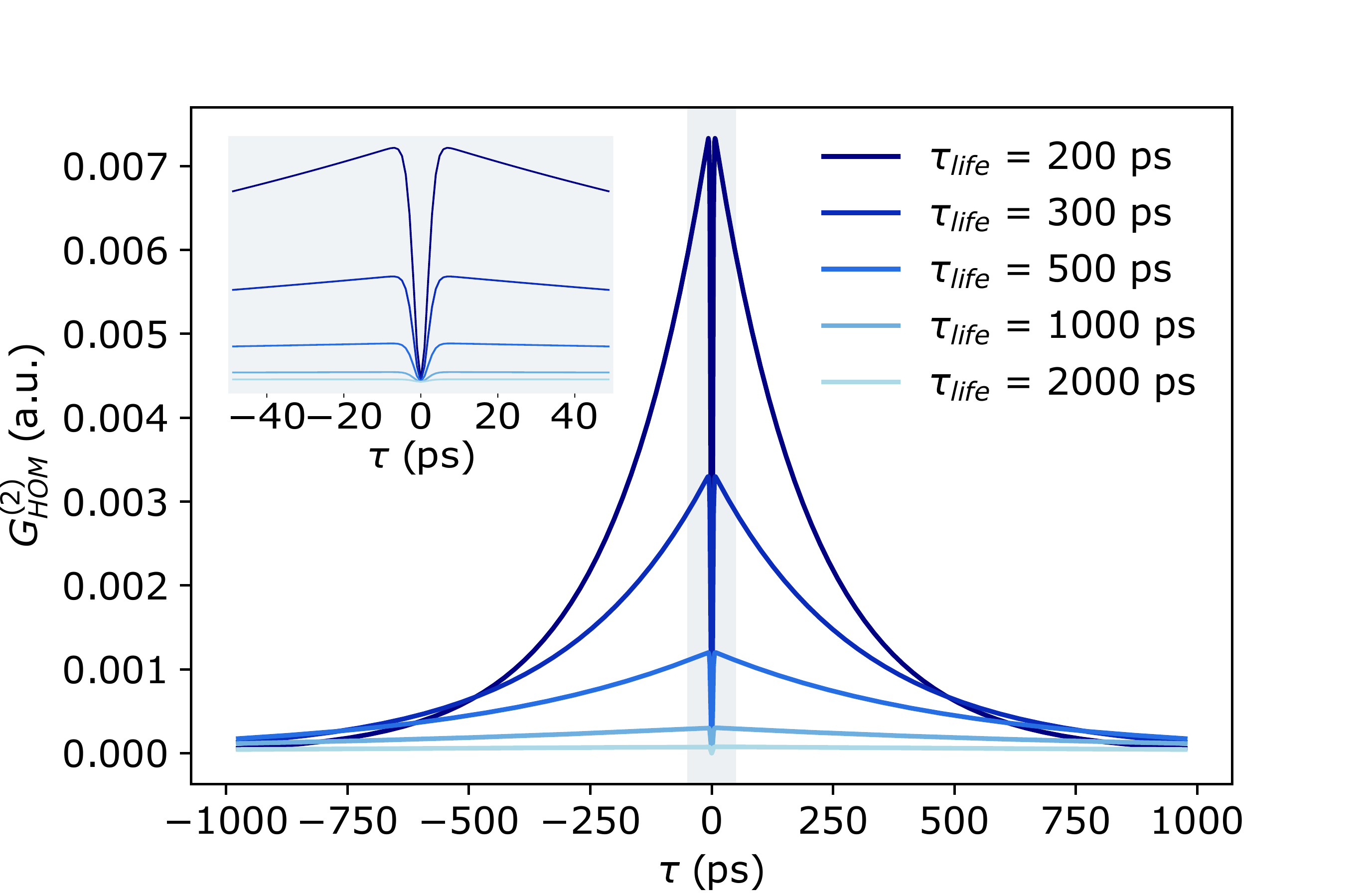}
	\caption{Time-resolved degree of HOM coherence $G^{(2)}_{HOM}(\tau)$ as a function of detection time difference $\tau$ in SI units. Identical emitters with lifetime $\tau_{life}$ and a fixed pulse width of $\tau_{pulse} = 5$ ps are assumed. Values are chosen to facilitate direct comparison to experimental data from ref. \cite{volcano1}. The inset depicts the grey shaded central region around $\tau=0$.} \label{fig:volcano}
\end{figure}

\noindent We now numerically calculate $G^{(2)}_{HOM}(\tau)$ functions for different system parameters and compare our calculations with typical experimental findings. In ref. \cite{volcano1} the HOM indistinguishability from single photons generated by single GaAs quantum dots with $\tau_{life} \approx 200$ ps was measured using 16 ps time bins. Setting a fixed pulse width of $\tau_{pulse} = 5$ ps as used in these experiments, we apply eqn. \ref{eq:g2HOM_timeresolved} to extend the findings to different emitter lifetimes. Figure \ref{fig:volcano} shows typical results for various lifetimes $\tau_{life}$ of identical emitters. The inset on the figure shows a zoom-in to the grey shaded region around $\tau = 0$ in the main plot. For $\abs{\tau} \gg \tau_{life}$, all correlations vanish since the probability of emitting photons decreases exponentially within the lifetime of the emitter. Thus, most of the correlations occur in a central region of $-3\tau_{life} < \tau < 3\tau_{life}$. The symmetry of $G^{(2)}_{HOM}(\tau)$ reflects the equivalent role of the two detectors.  
The origin of the non-zero correlations even for identical sources is a finite re-excitation probability during the excitation pulse \cite{reexc_qd}. Here, if the driven system emits a photon while still being addressed by the laser, there is a finite probability of re-excitation, and the emission of a second photon during the same excitation cycle. As the ratio $\tau_{pulse}/\tau_{life}$ decreases, this probability becomes gradually smaller as expected. Figure \ref{fig:volcano} confirms this behavior.

For $\abs{\tau}$ close to zero there is a rapid reduction of coincidences since the presence of a photon precludes the TLS being in the excited state. Thus, re-excitation is required resulting in a 'volcano-like' dip in the time-resolved degree of HOM coherence. As can be seen in the inset in figure \ref{fig:volcano}, $G^{(2)}_{HOM}(\tau)$ decreases from its maximum value at $\abs{\tau} \approx 7$ ps to reach zero at exactly $\tau=0$. Remarkably, the absence of coincidences at $\tau=0$ is found for single-photon wavepackets irrespective of their relative lengths and frequencies \cite{rempe}.

For our numerical simulations we assumed ideal experimental conditions in order to focus exclusively on the impact of the characteristic properties of the quantum emitters on two-photon interference properties. However, the finite time resolution in real experiments can obfuscate some of the features discussed here. 
For example, as a consequence of the finite detector temporal resolution the value for exactly $\tau=0$ is rarely measured since $G^{(2)}_{HOM}(\tau)$ is averaged over a finite interval around the origin. Thus, the central dip in figure \ref{fig:volcano} may not be observed in experiments. 
To quantify indistinguishability, it is not necessary to know the distribution of correlations with respect to $\tau$, but only the correlations summed over a specific time bin. These are connected to the probability of photons exiting at the different output ports and thus provide a measure for indistinguishability. Mathematically, we integrate the time-resolved degree of HOM coherence over a range of $\tau$ \cite{simulationpaper} and define the quantity $G^{(2)}_{HOM}(0)$ to be the \textit{pulse-wise} degree of HOM coherence, i.e. 

\begin{eqnarray}
G^{(2)}_{HOM}(0) &\equiv& \int_{0}^{\infty} \int_{-\infty}^{\infty} dt d\tau G^{(2)}_{HOM}(t,\tau) . \label{g2HOM_pulsed0}
\end{eqnarray}

\noindent This quantity corresponds to the total probability of having detection events at both detectors after exciting the two emitters with respective single pulses. A minimum value of 0 indicates perfectly indistinguishable single photons, which always exit on the same output port. A value of 0.5 is reached for two fully distinguishable single photons, meaning that the photons exit together in half of the cases and in the other half of the cases leave at different output ports (fully classical behavior). Any value smaller than 0.5 is non-classical and thus a signature of having at least partially indistinguishable photons. For independent input fields, values greater than 0.5 can only be obtained as a consequence of multi-photon emission.

The complete expression for the the pulse-wise degree of HOM coherence is

\begin{eqnarray}
&&G^{(2)}_{HOM}(0) = \frac{1}{4}( 
\int_{0}^{\infty} \int_{-\infty}^{\infty} dt d\tau (G^{(2)}_{11}(t, \tau) + G^{(2)}_{22}(t, \tau)) \nonumber \\ 
&&+  \int_{0}^{\infty} \int_{-\infty}^{\infty} dt d\tau (N_{1}(t)\cdot N_2(t+\tau) + N_2(t)\cdot N_1(t+\tau)) \nonumber \\ 
&&- \int_{0}^{\infty} \int_{-\infty}^{\infty} dt d\tau~ 2 \Re{G^{(1)}_{11}(t,\tau)^* \cdot G^{(1)}_{22}(t,\tau)} ) .\label{eq:g2HOM_pulsed}
\end{eqnarray}	

\noindent Substituting field operators with TLS ladder operators $\hat{\sigma_i}^{(\dagger)}$ and the decay rates $\gamma_i$ (see Appendix \ref{sec:app_opsubst}), photon indistinguishability can be expressed in terms of three different correlators. The subscript $i \in \{1,2\}$ denotes the respective emitter:

\begin{eqnarray}
G^{(2)}_{ii}(t,\tau) &=& \gamma_i^2 \expval*{\hat{\sigma}_i^{\dagger}(t)\hat{\sigma}_i^{\dagger}(t+\tau)\hat{\sigma}_i(t+\tau)\hat{\sigma}_i(t)} \label{eq:g2HOM_pulsed1}\\
N_i(t) &=& \gamma_i \expval*{\hat{\sigma_i}^{\dagger}(t) \hat{\sigma}_i(t)} \label{eq:g2HOM_pulsed2} \\
G^{(1)}_{ii}(t,\tau) &=& \gamma_i \expval*{\hat{\sigma}_i^{\dagger}(t) \hat{\sigma}_i(t + \tau)} . \label{eq:g2HOM_pulsed3}
\end{eqnarray}

\noindent The first line in eqn. \ref{eq:g2HOM_pulsed} is a sum of second-order auto-correlation functions of the two input fields. This term reflects single photon purity and thus accounts for possible multi-photon emission. The second line depends on the individual intensities and yields (not including the prefactor of 1/4) a constant value of $2$ if eqn. \ref{eq:onephoton} is satisfied \cite{simulationpaper}. This means that two-photon interference properties are fully governed by the third line, which is a product of field correlation functions of the two systems. If first-order coherence properties are similar in both input fields, its value becomes larger and in the case of indistinguishable photons exactly cancels the second line. Without the prefactor, this term is often referred to as the visibility V, such that $G^{(2)}_{HOM}(0) \gtrsim \frac{1}{2}(1-V)$.



Using eqn. \ref{eq:g2HOM_pulsed}, distinguishability due to different linear polarization angles can be accounted for by decomposing the ladder operators into orthogonally polarized components expressed by cosine and sine terms \cite{citationkai}. With a relative angle $\phi$ between the polarization directions of the two photons, eqn. \ref{eq:g2HOM_pulsed} is modified to account for polarization mismatch via the substitution:

\begin{eqnarray}
&& \Re{G^{(1)}_{11}(t,\tau)^* \cdot G^{(1)}_{22}(t,\tau)} \rightarrow \nonumber \\
&& \cos^2{(\phi)}  \Re{G^{(1)}_{11}(t,\tau)^* \cdot G^{(1)}_{22}(t,\tau)}. \label{eq:g2HOM_deltatau_pol}
\end{eqnarray}	

\noindent The cosine factor has no influence for parallel polarizations, while it leads to a vanishing interference term for orthogonal polarizations. Following a procedure frequently applied in experiments, where polarization filters and $\lambda$/2-plates are included in the HOM setup, we normalize $G^{(2)}_{HOM}(0)$ using cross-polarization \cite{polnorm1}. By considering the ratio of coincidences observed for parallel and orthogonal polarizations, a characterization of photon indistinguishability through the pulse-wise degree of HOM coherence can be maintained independent of incident photon flux.  This appraoch is also valid if less than one photon is emitted per pulse on average. Having identical coincidences in both polarization configurations indicates fully distinguishable photons. Observing fewer coincidences for parallel polarizations indicates that the photons are at least partially indistinguishable. The minimum value of zero uniquely corresponds to fully indistinguishable single photons. In order to bound the values to an interval [0,$\sim$0.5] in accordance to the non-normalized case, we additionally multiply the correlation-ratio with a factor of 1/2. We thus arrive at the \textit{polarization normalization factor} $\mathcal{N}_p $:

\begin{eqnarray}\label{eq:norm3}
&& \mathcal{N}_p = \frac{1}{2}( 
\int_0^{\infty} \int_{-\infty}^{\infty} dt d\tau (G^{(2)}_{1}(t, \tau) + G^{(2)}_{2}(t, \tau)) \\ 
&& + \int_0^{\infty} \int_{-\infty}^{\infty} dt d\tau (N_1(t)\cdot N_2(t+\tau) + N_2(t)\cdot N_1(t+\tau)). \nonumber 
\end{eqnarray}

\noindent In this work, we will exclusively use eqn. \ref{eq:norm3} for normalization (for further discussion on alternative normalization methods see Appendix \ref{sec:norm_app}). We thus define

\begin{eqnarray}
g^{(2)}_{HOM}(0) \equiv G^{(2)}_{HOM}(0)/\mathcal{N}_p  \label{eq:normeddegofcoh}
\end{eqnarray}

\noindent with $G^{(2)}_{HOM}(0)$ given by eqn. \ref{eq:g2HOM_pulsed}, $\mathcal{N}_p$ given by eqn. \ref{eq:norm3} and the lowercase g indicating that polarization normalization is applied.
In most cases considered, normalization will only have marginal influence on the results and can, in principle, be omitted. However, there are cases where an interpretation of $G^{(2)}_{HOM}$ as photon indistinguishability is not possible without using appropriate normalization (as may be the case for strong laser detuning, dephasing, or transmission losses in experiments).


\section{Influence of Emitter Properties}
\indent We continue to apply our methods to the case of $g^{(2)}_{HOM}(0)$ arising from two dissimilar emitters with a mutual spectral detuning $\Delta \omega$. An explicit incorporation of spectral detuning to eqn. \ref{eq:g2HOM_pulsed} can be found in Appendix \ref{sec:specdet}. We treat one of the two emitters as having a fixed decay rate $\gamma_1$, while the decay rate of the other is variable, represented by $\gamma_2$. By continuously varying $\gamma_2$, we tune the ratio $\gamma_2/\gamma_1$ to explore the influence of decay rate mismatches between the emitters, for any given spectral detuning $\Delta \omega$. The pulse width is fixed with respect to $\gamma_1$. It is chosen such that $\gamma_1 \tau_{pulse} \approx 0.026$, which yields a degree of second-order coherence of $g^{(2)}(0) \approx 0.008$ for similar sources. This value is motivated by the typical pulse duration used for quantum control experiments with III-V QD emitters.  Here, one typically uses a Ti:Sapphire laser delivering $\tau_{pulse}=10$ ps \cite{ti_sapph} duration pulses to excite a QD having $\tau_{life}=390$ ps \cite{390ps_qd}. Unless stated otherwise, these parameters are used for all simulation results presented below.

\begin{figure}
	\includegraphics[width=0.6\linewidth]{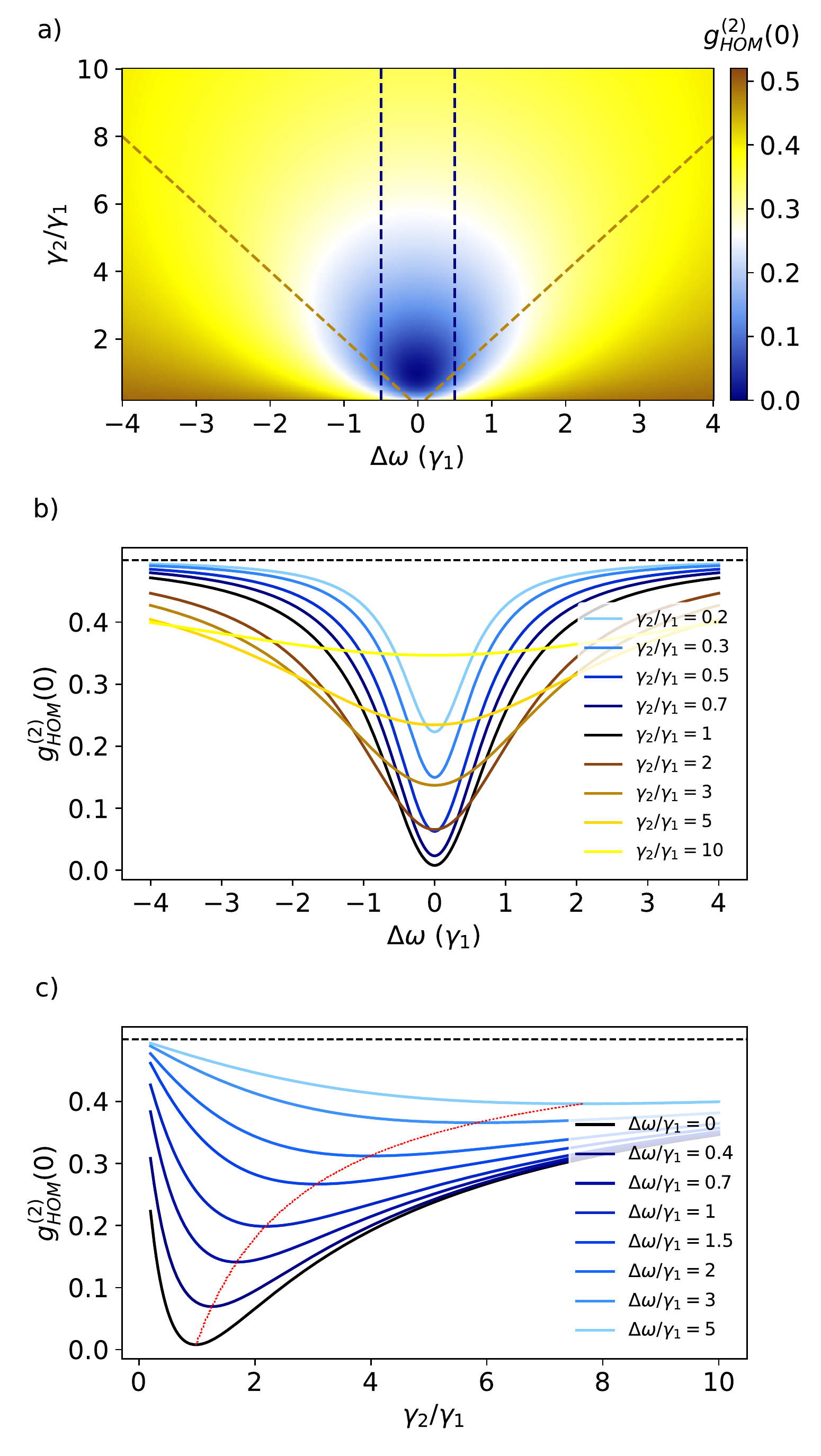}
	\caption{Influence of dissimilar decay rates $\gamma_2/\gamma_1$ and spectral detuning $\Delta \omega$ on photon indistinguishability. Decay rate $\gamma_1$ of emitter 1 is kept constant while tuning through the range of $\gamma_2 \in [0.2,10] \gamma_1$. (a) Pulse-wise degree of HOM coherence $g^{(2)}_{HOM}(0)$ as a function of spectral detuning $\Delta \omega$ and decay rate-ratio $\gamma_2/\gamma_1$. The dashed red and green lines represent the natural linewidths of emitters 1 and 2, respectively. (b) Horizontal cuts showing $g^{(2)}_{HOM}(0)$ as a function of $\Delta \omega$ for different $\gamma_2/\gamma_1$. 
	(c) Vertical cuts showing $g^{(2)}_{HOM}(0)$ as a function of $\gamma_2/\gamma_1$ for different $\Delta \omega$. The dashed red line serves as a guide to the eye to track minima of $g^{(2)}_{HOM}(0)$ for each $\Delta \omega$. The dashed horizontal line indicates the classical threshold of $0.5$.} \label{fig:gamma_omega}
\end{figure}

Figure \ref{fig:gamma_omega}a shows a false color image of the resulting pulse-wise degree of HOM coherence $g^{(2)}_{HOM}(0)$ as a function of spectral detuning and ratio of the decay rates. The dashed lines denote the bounds defined by the natural linewidths of the emitters.  The data presented in Figure \ref{fig:gamma_omega}a is characterized by a region around the origin for which $g^{(2)}_{HOM}(0)$ is minimized. Three physical phenomena connected to photon indistinguishability impact upon $g^{(2)}_{HOM}(0)$ when varying the decay rate-ratio: (i) re-excitation of the driven quantum emitters, (ii) the spatio-temporal overlap of the resulting photons on the beamsplitter and (iii) their natural linewidths. Figure \ref{fig:gamma_omega}b shows $g^{(2)}_{HOM}(0)$ as a function of spectral detuning for different $\gamma_2/\gamma_1$, corresponding to horizontal cuts in figure \ref{fig:gamma_omega}a. Moving away from $\gamma_2/\gamma_1 = 1$ results in reduced spatio-temporal overlap of the photons at the beamsplitter and thus increases $g^{(2)}_{HOM}(0)$. The overall minimum value of $g^{(2)}_{HOM}(0) = 0.008$ is found for identical emitters having a maximal spatio-temporal overlap. It is non-zero due to the finite re-excitation probability during the laser pulses \cite{reexc_qd}. Decreasing $\gamma_2$ reduces the re-excitation probability, but it also reduces the spatio-temporal overlap, resulting in a degredation of the overall indistinguishability. For higher $\gamma_2$, $g^{(2)}_{HOM}(0)$ becomes less susceptible to spectral detuning as a result of the increased natural linewidth and, therefore, increased spectral overlap of the photons. This can be more clearly seen in figure \ref{fig:gamma_omega}c, where $g^{(2)}_{HOM}(0)$ is plotted as a function of $\gamma_2/\gamma_1$ for various fixed spectral detunings, corresponding to vertical cuts in figure \ref{fig:gamma_omega}a. In the absence of spectral detuning, the optimum decay rate-ratio is $1$, indicating identical emitters. In the presence of finite spectral detuning, however, the minimum value of $g^{(2)}_{HOM}(0)$ is reached for $\gamma_2/\gamma_1 > 1$. The positions of the minima in $g^{(2)}_{HOM}(0)$ are indicated by the dashed red line in figure \ref{fig:gamma_omega}c. This shows that if the two quantum emitters are spectrally detuned by $\Delta \omega$, maximum indistinguishability is achieved when $\gamma_2 > \gamma_1$ and the two wavepackets have a maximum spectral overlap.  The increased natural linewidth associated with $\gamma_2/\gamma_1 > 1$ can overcompensate the detremental impact f increased re-excitation probability and reduced spatio-temporal overlap. Remarkably, this observation shows that there are situations where photon indistinguishability increases even if $\Delta \omega$ is increased.

\section{Processes limiting Two-Photon Interference visibility}

The framework we have developed thus far for quantifying photon indistinguishability takes into account intrinsic emitter mismatches bewteen the two quantum emitters, such as differences in lifetime and spectral detuning. We continue to also include extrinsic physical mechanisms that arise due to fluctuations of the environment of the quantum emitters, the methods used for quantum state preparation of experimental apparatus. 


We begin by exploring mismatches in photon arrival time at the beamsplitter.  
This can occur, for example, when the emitters are excited non-resonantly via a higher energy level and the population of the radiative state depends on incoherent relaxation processes, causing jitter in the photon arrival time at the beamsplitter \cite{nonclass_qdlight}. 


\noindent Mathematically, we account for a temporal delay into the expression for $g^{(2)}_{HOM}(0)$ (eqn. \ref{eq:g2HOM_pulsed}) by replacing the time variable $t$ for one of the two quantum emitters (denoted system 2) with a shifted variable $t-\delta \tau$ that accounts for the relative offset. We choose the minus sign by convention, indicating that positive temporal delays $\delta\tau >0$ correspond to later arrival times of the photon originating from system 2. By transforming the time variables we find that the first four terms in eqn. \ref{eq:g2HOM_pulsed} are not influenced by a temporal delay (see Appendix \ref{sec:tempdel_app}). However, for the final term in eqn. \ref{eq:g2HOM_pulsed} the four field operators each have different time dependencies and we include the temporal delay into this term explicitly.  This leads to the replacement:

\begin{eqnarray}
&&2 \Re{G^{(1)}_{11}(t,\tau)^* \cdot G^{(1)}_{22}(t,\tau)} \rightarrow \\
&&2 \Re{G^{(1)}_{11}(t,\tau)^* \cdot G^{(1)}_{22}(t-\delta\tau,\tau)}. 
\label{eq:g2HOM_deltatau} 
\end{eqnarray}	

\begin{figure}
	\centering
	\includegraphics[width=0.6\linewidth]{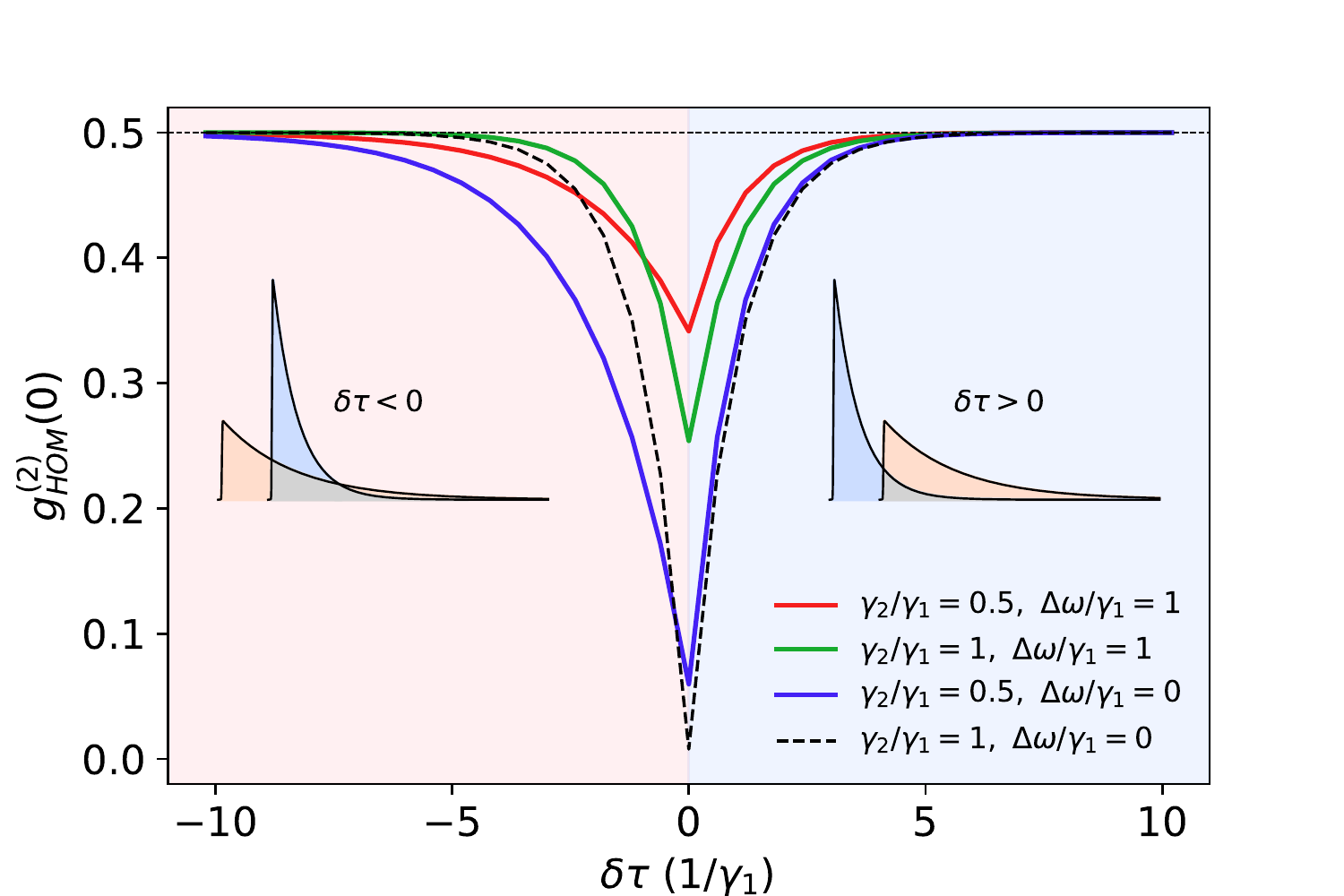}
	\caption{Pulse-wise degree of HOM coherence $g^{(2)}_{HOM}(0)$ as a function of the relative delay in photon arrival time $\delta \tau$. Positive values of $\delta \tau$ correspond to later arrival times of the photon from emitter 2. Instances are shown for identical emitters and for emitters differing in lifetime, frequency or both. The insets on the left and right illustrate the cases where the shorter or longer photon is delayed, respectively.} \label{fig:delay}
\end{figure}

\noindent Figure \ref{fig:delay} compares $g^{(2)}_{HOM}(0)$ as a function of $\delta \tau$ for identical photon wavepackets and for photons from emitters that differ in lifetime, frequency, or both. The minimum of $g^{(2)}_{HOM}(0)$ is always observed for $\delta \tau = 0$. For emitters with identical lifetimes and frequencies, introducing a time-delay leads to a symmetrical degradation of photon indistinguishability for both positive and negative values of $\delta \tau$. This effect can be readily understood as a manifestation of decreasing spatio-temporal overlap of the photon wavepackets at the beamsplitter as they are shifted with respect to each other in time. For $\abs{\delta \tau} = 1/\gamma_1$, $g^{(2)}_{HOM}(0)$ already $> 0.3$ and for $\abs{\delta \tau} = 3/\gamma_1$, with $g^{(2)}_{HOM}(0) = 0.48$ the classical threshold is almost reached. This is in agreement with the photonic probability density function, which drops to $1/e$ of its initial value within the lifetime of the emitter, suggesting that there is little to no overlap for even larger temporal mismatches.
  
For dissimilar decay rates such that $\gamma_2/\gamma_1=0.5$, the data presented in fig. \ref{fig:delay} reveal an asymmetry that can be explained by the fact that the photon originating from the longer-lived system is more spread out in the space-time domain. If the \textit{"longer"} photon wavepacket arrives earlier than the \textit{"shorter"} one, spatio-temporal overlap is reduced more slowly through the temporal delay than situation when the time ordering of the two wavepackets is interchanged (see inset on fig. \ref{fig:delay}). Experimentally, this means that measuring $g^{(2)}_{HOM}(0)$ versus $\delta \tau$ allows to classify whether the systems have equal or different lifetimes. By examining the width of the dip, a quantitative determination of the individual lifetimes can even be made.  For spectrally detuned photon wavepackets having equal spatio-temporal forms, figure \ref{fig:delay} shows that the behavior of $g^{(2)}_{HOM}(0)$ is qualitatively similar to the case of identical emitters with an overall reduced indistinguishability. Considering spectral and lifetime mismatch together leads to a combination of both individual effects: overall reduced indistinguishability with an asymmetric HOM dip as a function of $\tau$.


\noindent We continue to explore the impact of pure dephasing \cite{puredephexp2} on $g^{(2)}_{HOM}(0)$. Deoherence can either be caused by population decay or pure-dephasing.  Population decay arising, for example, by spontaneous emission has the inevitable side effect of causing coherence decay with half the population decay rate. Alternatively, a decay of the off-diagonal elements of the photon density matrix (pure dephasing) leaves the populations unaffected. The pure dephasing rate $\gamma_{deph}$ can be inferred from the $T_1$ lifetime and $T_2$ coherence times, which are the frequently used timescales in literature to characterize and compare the dynamics of quantum systems \cite{foxT1T2}. It generally holds that \cite{skinner1986puredephasing}

\begin{eqnarray}
\frac{1}{T_2} = \frac{1}{2 T_1} + \gamma_{deph} \label{eq:T1T2Tdeph}
\end{eqnarray}

\noindent where $\gamma_{deph}$ is the pure dephasing rate ($\neq 1/T_2$. Since we only consider spontaneous emission as a mechanism resulting in population decay, $T_1 = \tau_{life} = 1/\gamma$ and we use eqn. \ref{eq:T1T2Tdeph} to determine the pure dephasing collapse operator from the quantities $T_1$ and $T_2$ (see Appendix \ref{sec:puredeph_app}). 

\begin{figure}[H]
	\centering
	\includegraphics[width=0.6\linewidth]{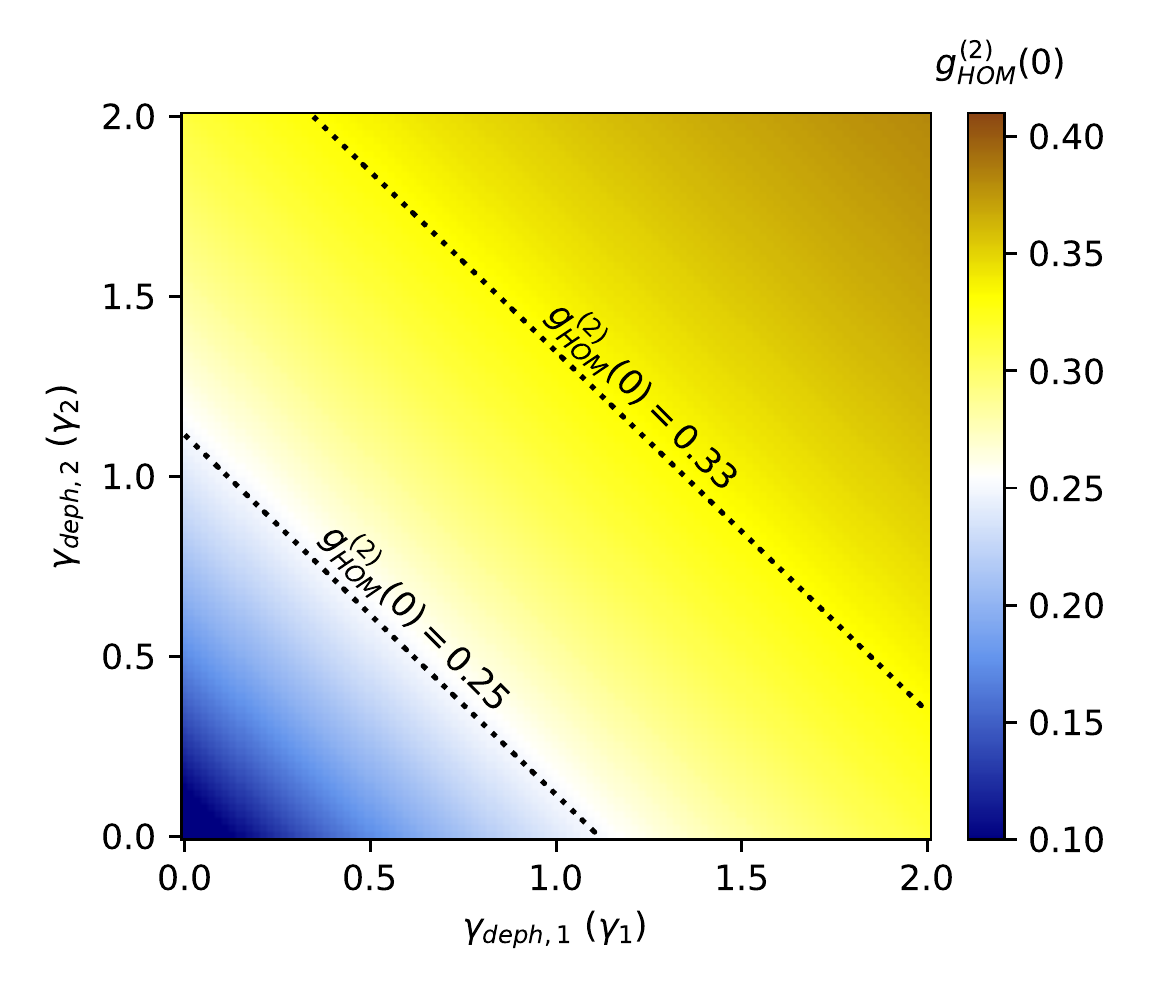}
	\caption{Pulse-wise degree of HOM coherence $g^{(2)}_{HOM}(0)$ as a function of individual dephasing rates $\gamma_{deph,1}$ and $\gamma_{deph,2}$ for emitters with different decay rates $\gamma_2=2\gamma_1$, but equal frequencies $\Delta\omega=0$. Note that for contrast enhancement the color bar is rescaled compared to figure \ref{fig:gamma_omega}a.} \label{fig:dephasingheatmap}
\end{figure}

Considering emitters having decay rates $\gamma_2=2\gamma_1$ and equal emission frequencies, fig. \ref{fig:dephasingheatmap} shows the pulse-wise degree of HOM coherence $g^{(2)}_{HOM}(0)$ as a function of their pure dephasing rates $\gamma_{deph,1}$ and $\gamma_{deph,2}$. Notably, we find that $g^{(2)}_{HOM}(0)$ only depends on the sum of dephasing rates $\gamma_{deph,12} = \gamma_{deph,1}+\gamma_{deph,2}$ and not explicitly on their individual values, a result that holds true independently of $\gamma_2/\gamma_1$. To achieve $g^{(2)}_{HOM}(0)<0.25$, we must have $\gamma_{deph,12} <1.1 \gamma_1$ and for $g^{(2)}_{HOM}(0)<0.33$ the sum of the dephasing rates must not exceed $2.3 \gamma_1$ (see dotted lines on the figure). Since only relative phase fluctuations between the photon wavepackets determine the two-photon interference behavior, it does not matter which emitter is subject to pure dephasing. In the following, we consider a model where pure dephasing is only present in system 1, characterized by a rate $\gamma_{deph,1}$. The generalization to the case of dephasing in both systems thus emerges naturally by replacing $\gamma_1$ by $\gamma_{12}$.

\begin{figure}
	\includegraphics[width=0.6\linewidth]{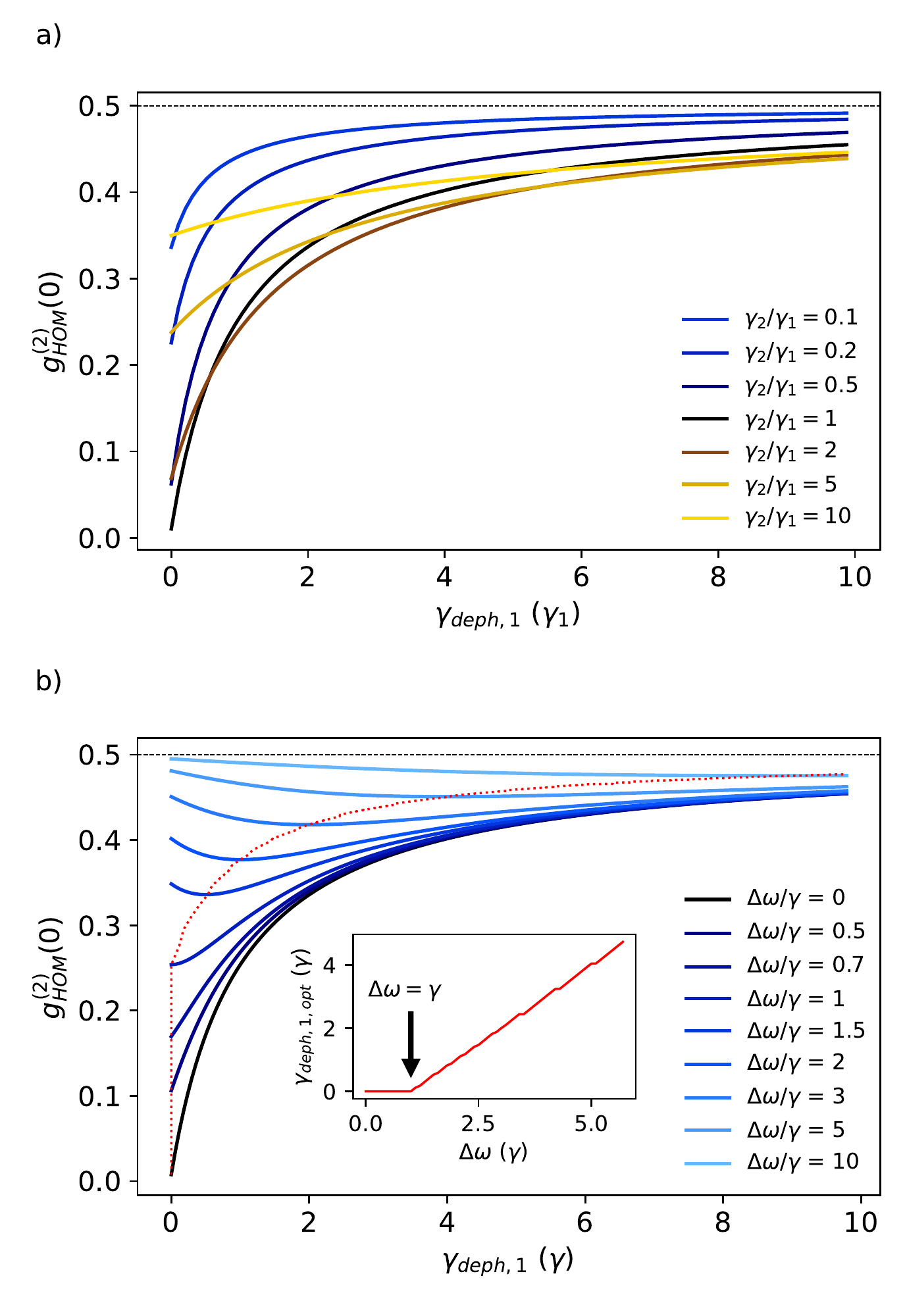}
	\caption{Pure dephasing in one system. Shown is the pulse-wise degree of HOM coherence $g^{(2)}_{HOM}(0)$ as a function of dephasing rate $\gamma_{deph,1}$ in system 1, for (a) zero spectral detuning $\Delta \omega=0$, and (b) identical decay rates $\gamma_2=\gamma_1$. Minima of $g^{(2)}_{HOM}(0)$ for each $\Delta \omega$ are indicated by a dashed line. The inset shows the dephasing rate $\gamma_{deph,1,opt}$, which yields minimal $g^{(2)}_{HOM}(0)$ for given $\Delta \omega$.} \label{fig:dephasing}
\end{figure}

Figure \ref{fig:dephasing} shows the effect on $g^{(2)}_{HOM}(0)$ of varying the natural linewidth and spectral detuning, in the presence of varying degrees of pure dephasing. 
Figure \ref{fig:dephasing}a illustrates the case for zero spectral detuning $\Delta \omega=0$, but variable decay rate-ratio $\gamma_2/\gamma_1$, where $\gamma_1$ is kept fixed. For similar sources, $g^{(2)}_{HOM}(0)$ increases from $<0.01$ to $0.38$ upon tuning the dephasing rate from $\gamma_{deph,1}=0$ to $3$.
Generally, the total increase in $g^{(2)}_{HOM}(0)$ depends on the linewidths of the emitters. In particular, the data presented in fig. \ref{fig:dephasing}a show that for a larger linewidth of emitter 2, $g^{(2)}_{HOM}(0)$ is more weakly affected by pure dephasing in system 1. For dephasing rates $\gamma_{deph,1}>5.5\gamma_1$, a decay rate-ratio $\gamma_2/\gamma_1=10$ even leads to more indistinguishable photons than when having similar sources. This shows that in the presence of strong dephasing, the effect of a broader linewidth can overcompensate for both reduced photon overlap and increased re-excitation. Although pure dephasing leads to a rapid degradation of indistinguishability, simulations show that the classical threshold is not exceeded up to values of $\gamma_{deph,1}>50\gamma_1$.

Figure \ref{fig:dephasing}b shows calculations of $g^{(2)}_{HOM}(0)$ as a function of $\gamma_{deph,1}$, assuming two emitters with equal decay rates $\gamma_1=\gamma_2\equiv \gamma$, but allowing for finite spectral detuning $\Delta \omega$. For $\Delta \omega \neq 0$, minimal $g^{(2)}_{HOM}(0)$ is generally not reached for $\gamma_{deph,1}=0$. For each $\Delta \omega$, there is a dephasing rate $\gamma_{deph,1,opt}$ that gives the smallest $g^{(2)}_{HOM}(0)$, indicated by the dashed red line on the figure. The value of $\gamma_{deph,1,opt}$ is presented as a function of $\Delta \omega$ in the inset in fig. \ref{fig:dephasing}b. Up to a spectral detuning that matches the natural linewidth, highest photon indistinguishability is achieved in the absence of pure dephasing, but for spectral detunings exceeding the natural linewidth, we find an approximately linear increase in $\gamma_{deph,1,opt}$ with $\Delta \omega$. Qualitatively, this can be understood by noting that pure dephasing introduces random phase shifts to the single photon wavepackets that interfere at the beamsplitter. In the absence of spectral detuning, this leads to a steady degradation of indistinguishability as the relative phase becomes increasingly randomized. However, in the presence of spectral detuning exceeding the natural indeterminacy of the emitters, these random phase shifts result in an occasional rephasing of the phase drift that stems from the frequency difference. In this way, phase randomization can partially counteract the frequency mismatch. However, this effect is typically too weak to be observed in experiments. For $\Delta \omega= 1.5 \gamma$ the gain arising from additional dephasing optimally leads to a reduction of $g^{(2)}_{HOM}(0)$ from $0.35$ to $0.34$. Comparing $g^{(2)}_{HOM}(0)$ for larger spectral detunings with and without dephasing present, there can be improvements in $g^{(2)}_{HOM}(0)$ of up to $8\%$ when having 'optimal' dephasing, but this only applies to conditions close to the classical threshold under which one would typically not perform two-photon interference experiments.

To validate our calculations we compare the predictions of our model with measurements reported in the literature. Ref. \cite{benyoucef} performed HOM measurements on two independent GaAs quantum dots from the same sample. HOM interference could not be observed in these experiments, an observation that was attributed to rapid dephasing. Using the measured $T_1$ and $T_2$ times in ref. \cite{benyoucef} and simulating the behavior under otherwise ideal conditions shows that the best achievable $g^{(2)}_{HOM}(0)$ is $0.42$. This clearly confirms pure dephasing to be the predominant source of indistinguishability degradation in this experiment, concurring with the conclusion of the authors and underscoring the necessity of considering it in theoretical descriptions. 


\noindent We then explore the impact of spectral wandering on the HOM interference of solid-state quantum emitters. Spectral wandering describes the impact of a noisy environment that randomizes the emission energy over timescales far larger than the radiative lifetime. This phenomenon is frequently encountered in solid state systems and is, for example, caused by a fluctuating charge environment that gives rise to electric field noise and, hence, frequency shifts of the quantum emitter due to the DC Stark effect \cite{specwan1,specwan2}. 
This leads to a probabilistic emission within a range of frequencies around the center frequency $\omega_0$ of the emitter with a spectrum that is sensitive to the details of noise spectrum. Each individual emission occurs with the natural linewidth $\gamma$ of the emitter, but averaging successive emissions over time leads to a broadened linewidth \cite{specwan_linewidth}. 

We have seen above that the value of $g^{(2)}_{HOM}(0)$ explicitly depends on $\Delta \omega$. To connect spectral wandering in one or in both emitters to photon indistinguishability, we average $g^{(2)}_{HOM}(0,\Delta \omega)$ over all possible spectral detunings, weighted by their probability of occurrence, $p(\Delta \omega)$: 

\begin{eqnarray}
g^{(2)}_{HOM}(0) = \int_{-\infty}^{\infty} d\Delta \omega~ p(\Delta \omega) g^{(2)}_{HOM}(0, \Delta \omega). \label{eq:specdet}
\end{eqnarray}

To evaluate eqn. \ref{eq:specdet}, the probability distribution of spectral detunings between the two systems, $p(\Delta \omega)$ is required. This can be found from the distributions of emission frequencies $p(\omega_i)$ of the individual systems through a variable transformation of the combined probability distribution function, i.e.:  

\begin{eqnarray}
p(\Delta \omega) &=& \int_{-\infty}^{\infty} \int_{-\infty}^{\infty} d\omega_1 d\omega_2 \delta(\Delta \omega -(\omega_2 - \omega_1))  p(\omega_1, \omega_2). \label{eq:transfpdf}
\end{eqnarray}

\noindent Here, we assume the frequency distributions of both emitters are independent, such that no correlations exist between frequency fluctuations in the different systems. This implies that $p(\omega_1, \omega_2) = p(\omega_1)\cdot p(\omega_2)$ and is clearly the case for dissimilar sources in separate samples. In accord with experimental studies of fluctuation dynamics in III-V quantum dots, we model the distribution of emission frequencies in each emitter i $\in \{1,2\}$ using a Gaussian distribution \cite{gauss_wan, motionalnarrowing}:

\begin{eqnarray}
p(\omega_i) = \frac{1}{\sqrt{2 \pi \sigma_i^2}} \exp{-\frac{1}{2 \sigma_i^2}(\omega_i-\omega_{0i})^2} . \label{eq:wandering0}
\end{eqnarray}

\noindent In this case, the normalized probability distribution function is fully determined by its center frequency $\omega_{0i}$ and variance $\sigma_i^2$. The latter is connected to the full width at half maximum (FWHM) via FWHM$_i = \sqrt{8 \ln(2)) \sigma_i^2}$. In general, the frequency distributions of the two emitters have different widths and peak positions. Combining eqs. \ref{eq:transfpdf} and \ref{eq:wandering0} we find the distribution of spectral detunings as required for eq. \ref{eq:specdet} (see Appendix \ref{sec:specwan_app} for the explicit integration):

\begin{eqnarray}
p(\Delta \omega) &=& \frac{1}{\sqrt{2 \pi (\sigma_1^2+\sigma_2^2)}} \exp{-\frac{1}{2}\frac{(\Delta \omega -\Delta \omega_0)^2}{\sigma_1^2 + \sigma_2^2}}. \label{eq:spectralwandering}
\end{eqnarray}

\noindent We see that the variances $\sigma_i^2$ of the individual distributions are added, such that the width of the transformed distribution is

\begin{eqnarray}
\text{FWHM}_{\Delta \omega} = \sqrt{\text{FWHM}_1^2 +\text{FWHM}_2^2} . \label{eq:FWHM_add}
\end{eqnarray}

\noindent Thus, the influence of spectral wandering depends on the center frequencies and the sum of the variances of both distributions, rather than the widths of the individual distributions themselves. Note that for identical distributions according to eqn. \ref{eq:FWHM_add}, the width is increased by a factor $\sqrt{2}$. In the case where only one system experiences spectral wandering, the probability distribution function of $\Delta \omega$ has the same form as the individual frequency distribution of the emitter.

\begin{figure}
	\includegraphics[width=0.6\linewidth]{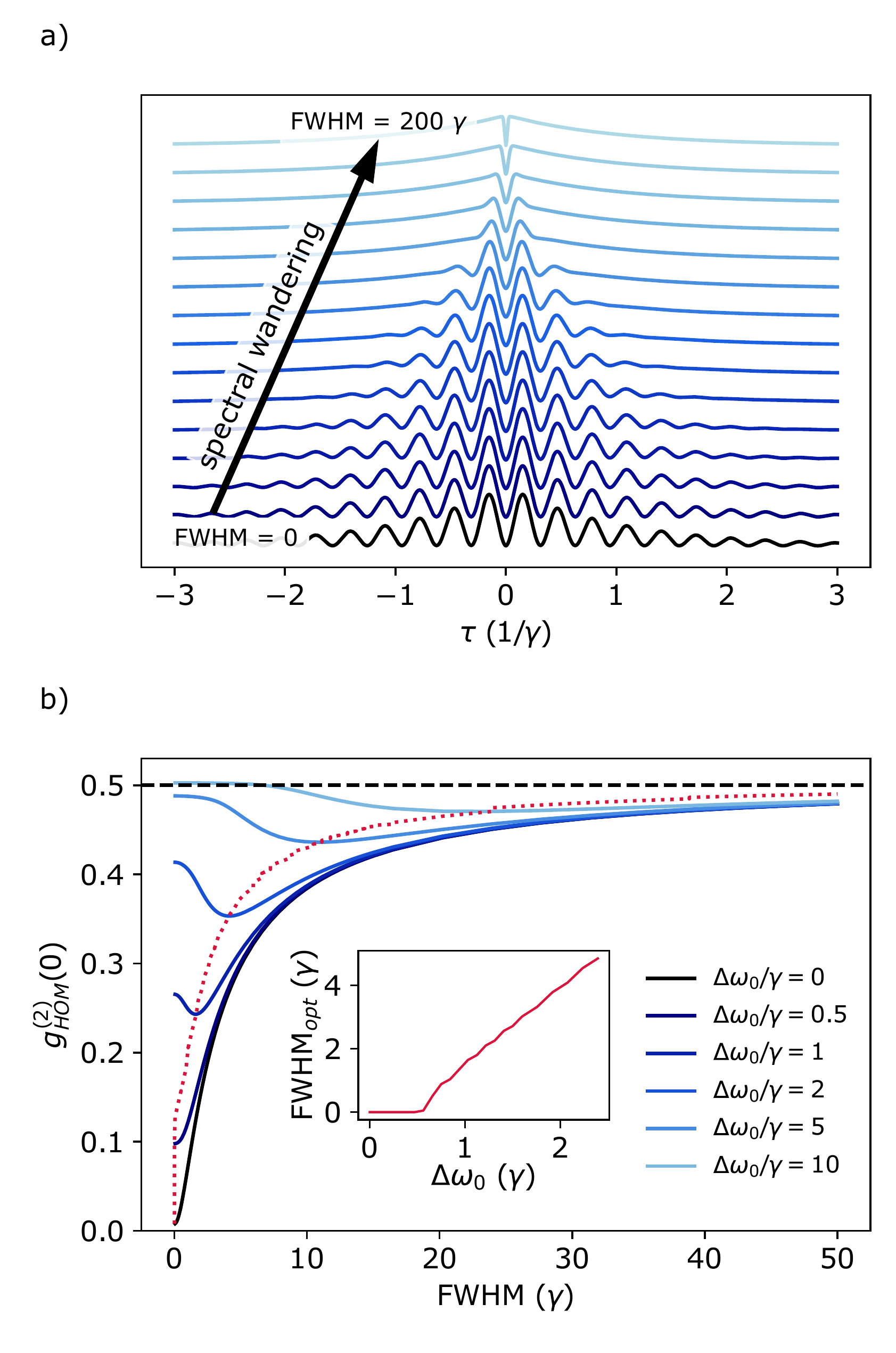}
	\caption{(a) Degradation of quantum beats due to the fluctuating noise environment of the quantum emitters. The time-resolved degree of HOM coherence $G^{(2)}_{HOM}(\tau)$ is shown for increasing full-width at half maximum FWHM of the frequency distribution of one emitter. We assume equal decay rates $\gamma_1 = \gamma_2 \equiv \gamma$ and a strong spectral detuning of $\Delta\omega_0 = 20 \gamma$. (b) Pulse-wise degree of HOM coherence $g^{(2)}_{HOM}(0)$ as a function of the width FWHM of the frequency distribution of emitter 2 for different spectral detunings $\Delta\omega_0$. We assume emitters with equal decay rates $\gamma$ and no spectral wandering in system 1. The dashed red line serves as a guide to the eye to indicate respective minima. The inset shows the width of the frequency distribution yielding minimal $g^{(2)}_{HOM}(0)$, denoted FWHM$_{opt}$, as a function of $\Delta\omega_0$.} \label{fig:specwan}
\end{figure}

The influence of noise is manifested not only in the overall photon indistinguishability, but also in the distribution of correlations. For noiseless environments, a quantum beat signal can be seen in the time-resolved degree of HOM coherence in the presence of spectral detuning. Theoretically, such a behavior was predicted by Legero et al \cite{qbeats_rempe} and has been observed experimentally with photons from atoms \cite{qbeat_rempe}, molecules \cite{qbeat_molecules} or quantum dots \cite{qbeat_qds}. Figure \ref{fig:specwan}a shows how such a quantum beat signal is influenced by gradually adding noise to the environment. The time-resolved degree of HOM coherence $G^{(2)}_{HOM}(\tau)$ is presented as a function of the time difference $\tau$ between detection event with and for increasing spectral wandering in emitter 1. We assume a spectral detuning of $\Delta \omega_0 = 20 \gamma$, such that we are in the regime of fully distinguishable photons with $g^{(2)}_{HOM}(0) \approx 0.5$. 

In the absence of noise, we observe a decaying cosinusoidal oscillation of the coincidence probability. Increasing noise in the environment leads to a probabilistic emission within a broader range of frequencies. Thus, we have to take a statistical average of $G^{(2)}_{HOM}(\tau)$ over the possible spectral detunings. Since the quantum beat frequency is determined by the absolute value of the spectral detuning, this corresponds to averaging over quantum beat signals having different frequencies. This leads to a successive smoothing of the beats for increased spectral wandering. As shown in fig. \ref{fig:specwan}a for  FWHM approaching 200$\gamma$, the oscillatory behavior vanishes, and besides the volcano-like dip, the coincidence probability steadily decreases toward larger $\abs{\tau}$.

We continue to consider two emitters with equal decay rates $\gamma$, whose center frequencies are detuned by $\Delta \omega_0=\omega_{02}-\omega_{01}$. Figure \ref{fig:specwan}b shows the pulse-wise degree of HOM coherence $g^{(2)}_{HOM}(0)$ as a function of the FWHM of the underlying Gaussian frequency distribution of emitter 2 subjected to a noisy environment. Emitter 1 is considered to be noise-free and various spectral detunings are compared (indicated by line color). For $\Delta \omega_0=0$, spectral wandering in the range of the natural linewidth leads to a steady increase of $g^{(2)}_{HOM}(0)$ from 0.008 to 0.07, reaching up to $g^{(2)}_{HOM}(0)=0.3$ for FWHM = 5$\gamma$. However, if the emitters are spectrally detuned, spectral wandering can result in them occasionally becoming resonant. For a range of widths of the frequency distribution this leads to an increase in indistinguishability as compared to the case without additional noise. Although the symmetry in the frequency distribution makes it equally likely for the frequencies to be further / less detuned, there can be an overall improvement in $g^{(2)}_{HOM}(0)$. This is a result of the highly non-linear dependence of $g^{(2)}_{HOM}(0)$ on $\Delta \omega$. In particular, we expect a strong degradation of $g^{(2)}_{HOM}(0)$ within a range of about $\Delta \omega=0-3\gamma$, but comparatively little effect for larger spectral detunings. If the FWHM is very small compared to $\Delta \omega_0$, the emission energies never become resonant, if the FWHM is very large, there is a dominant spectral detuning in the other direction. Thus, for any given $\Delta \omega_0$, there is a particular FWHM that yields a minimal $g^{(2)}_{HOM}(0)$, which we denote by FWHM$_{opt}$. The inset in figure \ref{fig:specwan}b shows the dependence of FWHM$_{opt}$ on $\Delta \omega_0$. Up to spectral detunings of $\Delta \omega_0= 0.5 \gamma$, where the natural linewidths of the emitters overlap each other's center frequency, there can be no improvement to the indistinguishability by adding noise. However, as the spectral detuning becomes larger, there is an optimal linewidth that leads to improved indistinguishability adhering to an approximately linear behavior. Although this effect is most likely too small to be explicitly exploited experimentally, it means that when facing spectrally detuned emitters, one can relax concerns about noisy environments up to a certain extent.

\section{Assessing hybrid combinations}

\noindent We now continue to utilize our model to explore hybrid quantum network architectures, in which different emitter combinations are used to generate indistinguishable photons. Emitter 1 is chosen to have a lifetime of either $\tau_{life}=250$ ps or $\tau_{life}=2$ ns, representative of GaAs quantum dots grown using droplet epitaxy \cite{qdl1,qdl2,qdl3}, or Purcell enhanced self-assembled InAs \cite{qdsk,qdsk2,qdsk3} quantum dots, respectively. The $\tau_{life}=2$ ns value is chosen to represent quantum emitters arising from e.g. atomic scale defects in 2D materials\cite{2dlifetimes} or color centers in diamond \cite{aharonovich2016solid}. The emission wavelength of the second emitter is assumed to be controllable, e.g. via DC Stark effect tuning such that it can be tuned precisely into resonance with emitter 1. For excitation, we consider resonant, coherent state preparation using a laser $\pi$-pulse having with a width of $\tau_{pulse}=10$ ps for each system.  Pure dephasing is not taken into account.

\begin{figure}[H]
	\centering
	\includegraphics[width=0.6\linewidth]{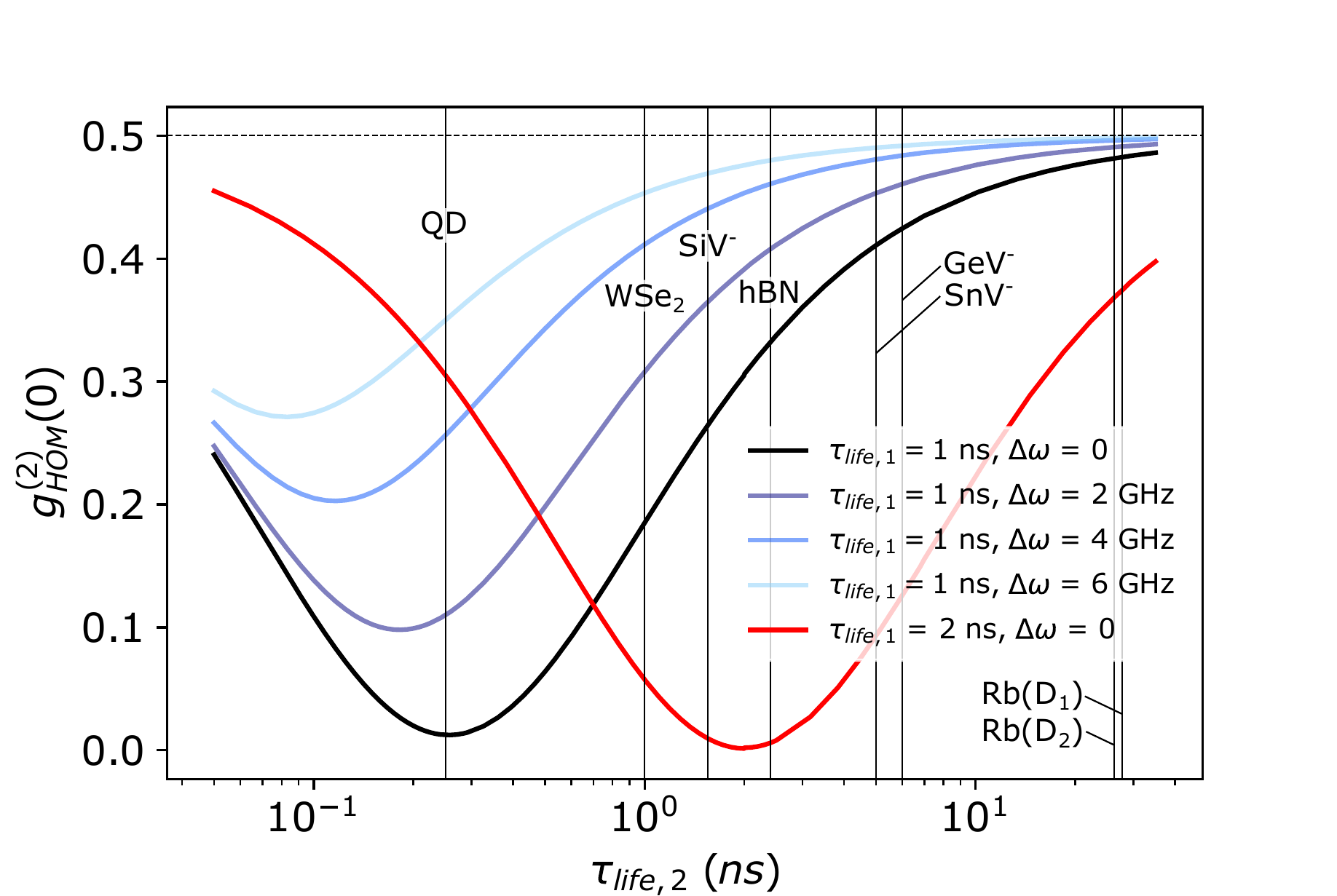}
	\caption{Comparison of different real systems paired with an emitter with lifetime $\tau_{life,1}=250$ ps, corresponding to a natural linewidth of $\Delta\omega=4$ GHz (black/blue), or $\tau_{life,1}=2$ ns (red). The pulse-wise degree of HOM coherence $g^{(2)}_{HOM}(0)$ is plotted as a function of the lifetime of the second emitter $\tau_{life,2}$, in the case of $\tau_{life,1}=250$ ps also for various spectral detunings $\Delta{\omega}$. Vertical lines on the figure denote the lifetimes of different real quantum emitters serving as potential candidates in quantum networks: WSe$_2$ \cite{wse2_lifetime}, hBN \cite{hbn_lifetime}, SiV$^-$ \cite{siv_lifetime}, SnV$^-$ single emitters \cite{snv_lifetime} and  GeV$^-$ in bulk diamonds \cite{gev_lifetime} and the D-lines of Rb-85 $\ce{^{85}Rb}$(D$_1$/D$_2$) \cite{rbd1_lifetime,rbd2_lifetime}.} \label{fig:real}
\end{figure}

\noindent Figure \ref{fig:real} shows the calculated value of $g^{(2)}_{HOM}(0)$ as a function of the lifetime of system 2, $\tau_{life,2}$, plotted on a logarithmic scale. For the case when $\tau_{life,1}= 250$ ps, we include spectral detunings from 0 to 6 GHz that may be encountered in typical experiments. Lifetimes of example systems potentially used in quantum networks are indicated by vertical lines. The plot compares quantum emitters induced in 2D materials via strain or point defects (WSe$_2$, hBN), color centers in bulk diamond (SiV$^-$, SnV$^-$, GeV$^-$) and atomic transitions (Rb(D$_1$), Rb(D$_2$)). Figure \ref{fig:real} serves as a reference that provides information on which component quantum emitters a quantum dot, 2D emitter or color center could readily be combined with in a hybrid quantum network.

With the quantum dot as emitter 1 (black and blue curves on fig. \ref{fig:real}) we find that a maximum lifetime of $\tau_{life,2}=4.5$ ns is possible for the second emitter when $\Delta\omega=0$ in order to not exceed a value of $g^{(2)}_{HOM}(0)=0.4$. Remarkably, we note that this is $\sim18\times$ larger than the quantum dot lifetime. In the presence of a spectral detuning of $\Delta\omega=4$ GHz, corresponding to the natural linewidth of the quantum dot, the maximum tolerable lifetime of emitter 2 needed to still achieve $g^{(2)}_{HOM}(0) \leq 0.4$ reduces to $\tau_{life,2}=0.9$ ns (3.6x the quantum dot lifetime). When pairing a quantum dot with an emitter of shorter lifetime than itself, the larger linewidth of the other emitter would reduce the impact of spectral detuning. But typical quantum dots have shorter lifetimes than other potential quantum emitters, such that their linewidth primarily determines the susceptibility to spectral detuning.
As shown by the red curve on fig. \ref{fig:real}, taking a 2D material or color center as emitter 1, we see that the high-indistinguishability regime where $g^{(2)}_{HOM}(0) < 0.1$ extends from $\tau_{life,2}\approx$ 0.8 to 5.2 ns. The use of these emitter types thus bridges the gap to hybrid systems involving higher-period vacancy centers with lower $\gamma$.

Overall, we can infer that theoretically, non-classical $g^{(2)}_{HOM}(0)$ could be achieved for quantum emitters with lifetimes that differ by more than a factor of $100\times$. 
The most promising combinations are QDs with SiV$^-$ centers in diamond or, in the future, possibly with 2D emitters. Combining quantum dots with atoms or ions, Purcell enhancement \cite{purcell_atom_gen} would be necessary to generate indistinguishable photons.

\section{Conclusion}

\noindent In summary, we have derived a general formalism that is capable of quantitatively characterizing two-photon interference from dissimilar sources subject to resonant pulsed quantum state preparation.  Our methods are valid for both time-resolved and pulse-wise integrated forms. We incorporated the key parameters that have an impact on indistinguishability: emitter decay rate, spectral detuning, temporal mismatch, pure dephasing and spectral wandering (see Appendix \ref{sec:fulleq} for summary of full mathematical incorporation). 

\begin{table}[H]
	\centering
	\begin{tabular}{@{} *5l @{}}    \toprule
		$g^{(2)}_{HOM}(0)$  & <0.1 &<0.2&<0.3  \\\midrule
		$\gamma_2/\gamma_1$    & ~2.5  & ~4.2  & ~7.3    \\ 
		$\abs{\Delta\omega_0}~(\gamma)$ & ~0.5  & ~0.8  & ~1.2 \\ 
		$\abs{\Delta\tau}~(1/\gamma)$  & ~0.3 & ~0.5 & ~1.0 \\
		$\gamma_{deph,12}~(\gamma)$  & ~0.2  & ~0.6  & ~1.4  \\
		$\text{FWHM}_{12}~(\gamma)$  & ~1.3  & ~2.7  & ~5.3  \\\bottomrule
		\hline
	\end{tabular}
	\caption{Maximal mismatches tolerated to achieve a pulse-wise degree of HOM coherence $g^{(2)}_{HOM}(0)$ below a given threshold. For otherwise ideal conditions, the individual influence of decay rate mismatch $\gamma_2/\gamma_1$, spectral detuning $\Delta\omega_0$, temporal delay $\Delta\tau$, pure dephasing rate in both systems and spectral wandering in both emitters is calculated.}\label{tab:values}
\end{table}

\noindent Table \ref{tab:values} compares the individual maximal offsets that could be tolerated to achieve $g^{(2)}_{HOM}(0)$ below a certain threshold for decay rate mismatch $\gamma_2/\gamma_1$, spectral detuning $\Delta\omega_0$, temporal delay $\Delta\tau$, combined pure dephasing rate in both systems $\gamma_{deph,12}=\gamma_{deph,1}+\gamma_{deph,2}$ and width of frequency mismatch distribution for Gaussian spectral wandering in both emitters $\text{FWHM}_{12}=(\text{FWHM}_1^2+\text{FWHM}_2^2)^{1/2}$.  We note that the precise offsets that could be tolerated in any real world quantum communication scenario would also be defined by the specific protocol that was being implemented. For each non-ideality, otherwise ideal conditions were considered in order to isolate the specific impact of each. We particularly note that spectral detuning and pure dephasing lead to a rapid degradation of indistinguishability. These are indeed often the main causes why photon coalescence is not observed in HOM experiments \cite{benyoucef, qtelimperfdots}. However, our simulations have shown that two different quantum emitters can be expected to exhibit measurable HOM quantum interference, even when their intrinsic properties differ.  Since there is a direct relation between HOM visibility and entanglement fidelity \cite{fidelity_visibility}, we believe that the framework presented in this work will be useful to benchmark future hybrid combinations for quantum networks based on two-photon interference.

\begin{acknowledgments}
\noindent We gratefully acknowledge financial support from the German Federal Ministry of Education and Research via Q.Link.X (16KIS0874), QR.X (16KISQ027) and the funding program Photonics Research Germany (Contract No. 13N14846), the European Union’s Horizon 2020 research and innovation program under Grant Agreements 820423 (S2QUIP) and 862035 (QLUSTER) and the Deutsche Forschungsgemeinschaft (DFG, German Research Foundation) via SQAM (FI947-5-1), DIP (FI947-6-1), and the Excellence Cluster MCQST (EXC-2111, 390814868). AJB acknowledges EPSRC grants EP/T017813/1, EP/T001062/1 and Marie Curie ITN project LasIonDef (GA n.956387). We also thank Rahul Trivedi for valuable discussions.
\end{acknowledgments}

\appendix

\section{Gaussian Excitation Pulse} \label{sec:gauss_ex}

\noindent A Gaussian pulse shape implies a time-dependence given by an electric field amplitude of the form 

\begin{eqnarray}
E_0 (t) = E_0(t_0) \cdot  e^{-\frac{1}{2}(\frac{t-t_0}{\sigma_{pulse}})^2},
\end{eqnarray}

\noindent where $E_0(t_0)$ is the maximum field amplitude, occurring at time $t_0$. The pulse parameter $\sigma_{pulse}$ is connected to the full width at half maximum of the pulse envelope (FWHM), which we define as the pulse width $\tau_{pulse}$:

\begin{eqnarray}
\tau_{pulse} \equiv \text{FWHM} = \sigma_{pulse} \cdot \sqrt{8 ln(2)} .
\end{eqnarray}

\noindent We also fix the pulse area to a value of $\pi$, resulting in an average of one photon emitted per pulse:

\begin{eqnarray}
\int \Omega(t) dt = \pi , \label{eq:pipulse}
\end{eqnarray}

\noindent with the Rabi frequency $\Omega$(t).

\section{Substituting field by TLS operators} \label{sec:app_opsubst}

In the absence of dephasing and for resonant excitation, relation \ref{eq:pipulse} is approximately equivalent to the normalization condition \cite{simulationpaper}

\begin{eqnarray}
\int dt \gamma \expval*{\hat{\sigma}^{\dagger}(t)\hat{\sigma}(t)} &=& 1 , \label{eq:onephoton}
\end{eqnarray}

\noindent where $\hat{\sigma}^{(\dagger)}$ is the TLS annihilation (creation) operator. Equation \ref{eq:onephoton} states that if a $\pi$-pulse is applied, one photon is emitted on average. We thus identify the integrand $\gamma \expval*{\hat{\sigma}^{\dagger}(t)\hat{\sigma}(t)}$ as the probability density of having a photon emitted at time $t$. This is the key step for incorporating the system dynamics into the correlation functions in eqn. \ref{eq:ghomttau}. Explicitly, this is done by expressing the input field operators $\hat{a}'^{(\dagger)}$ and $\hat{b}'^{(\dagger)}$ through the lowering and raising operators $\hat{\sigma}_1^{(\dagger)}$ and $\hat{\sigma}_2^{(\dagger)}$ of TLS 1 and 2, respectively:

\begin{eqnarray}
\hat{a}'^{(\dagger)} \rightarrow \sqrt{\gamma_1} ~ \hat{\sigma}_1^{(\dagger)} \label{eq:sysdyn1}\\
\hat{b}'^{(\dagger)} \rightarrow \sqrt{\gamma_2} ~ \hat{\sigma}_2^{(\dagger)} \label{eq:sysdyn2}.
\end{eqnarray}

\noindent In this way, the excitation of the TLS translates into an excitation of the field through its decay rate. Performing these replacements in eqns. \ref{eq:ghomttauterms1}-\ref{eq:ghomttauterms3}, we arrive at a modified version of eqn. \ref{eq:ghomttau} that exclusively depends on the dynamics of the emitters. To be consistent with the units of $1/s^2$, $G^{(2)}_{HOM}(t, \tau)$ now has to be interpreted as a correlation density for the detection times $t$ and $t+\tau$.

\section{Master Equation in Lindblad Form}

We calculate the correlators as defined in equations \ref{eq:ghomttauterms1}-\ref{eq:ghomttauterms3} by determining the time evolution of the Heisenberg operators. For the description of time evolution, we chose the master equation in Lindblad form, since it readily allows for the inclusion of dissipative, and thus non-unitary dynamics of the driven two-level systems \cite{preskill_lecture}:

\begin{widetext}
\begin{equation}
\pdv{t} \hat{\rho}(t) = {-\frac{i}{\hbar}[\hat{H}, \hat{\rho}]} + {\sum_{\mu=1}^{M}( \hat{L}_\mu \hat{\rho}(t)\hat{L}_\mu^{\dagger}-\frac{1}{2}\hat{L}_\mu^{\dagger}\hat{L}_\mu\hat{\rho}(t)-\frac{1}{2}\hat{\rho}(t)\hat{L}_\mu^{\dagger}\hat{L}_\mu)}.\label{eq:Lindblad_eq}
\end{equation}
\end{widetext}

\noindent The first term in eqn. \ref{eq:Lindblad_eq} corresponds to the unitary time evolution governed by the von-Neumann equation and involves the Hamiltonian $\hat{H}$ describing the light-matter interaction of the driven two-level system. The summation term, known as the dissipator $\mathcal{D}(\hat{\rho})$, accounts for any additional dissipative effects due to interactions with the environment. Each individual term in the sum in $\mathcal{D}$ is defined through its corresponding Lindblad (or collapse) operator $\hat{L}_\mu$ that can represent mechanisms such as spontaneous emission or pure dephasing. Note that in deriving equation \ref{eq:Lindblad_eq}, the so-called Markov approximation is used, which requires a separation of time scales on which the environment can store and retransfer information to the system from time scales inherent to the dynamics of interest. When considering processes involving phonons, such as for example the electron-phonon interaction in semiconductors, environment memory times may be on the same order of magnitude as light-matter interactions \cite{phonons_qo}. Therefore, an explicit inclusion of phonons within the framework of equation \ref{eq:Lindblad_eq} is generally not possible, and other approaches have to be employed \cite{phononsreview}. However, by using experimentally determined emitter decay rates, the influence of phonons on the resonant emission is implicitly accounted for. 

\section{Spontaneous Emission}
\label{sec:spontem}

\noindent Spontaneous emission arises because a two-level emitter inevitably interacts with vacuum modes of the electromagnetic field \cite{griffiths2018introduction} and causes the excited state to have a characteristic lifetime $\tau_{life}$, after which it decays to the ground state by emitting a photon. The finite lifetime of the excited state also leads to a spectral width of the emission line, as can be inferred from the energy-time uncertainty relation

\begin{eqnarray}
\Delta E \cdot \Delta t \geq \hbar &\Leftrightarrow&  \Delta \omega \geq \gamma. \label{eq:linewidth}
\end{eqnarray}

\noindent So the emission line a TLS is not sharply defined, but indeterminate in a small area around its center. The lower bound for this range $\Delta \omega$ is given by its decay rate $\gamma=1/\tau_{life}$ and is called the natural linewidth.

We incorporate different emitter lifetimes into the description of the system dynamics by finding the corresponding collapse operator in the Lindblad equation (eqn. \ref{eq:Lindblad_eq}). From the time evolution of a TLS density matrix under spontaneous emission \cite{steck_optbloch}:

\begin{eqnarray}
\pdv{t} \rho_{gg}(t) &=& \gamma \rho_{ee}(t), \label{eq:optbloch_00}\\ 
\pdv{t} \rho_{ge}(t) &=& - \frac{\gamma}{2} \rho_{ge}(t), \label{eq:optbloch_01}\\ 
\pdv{t} \rho_{eg}(t) &=& - \frac{\gamma}{2} \rho_{eg}(t), \label{eq:optbloch_10}\\ 
\pdv{t} \rho_{ee}(t) &=& -\gamma \rho_{ee}(t) \label{eq:optbloch_11} ,
\end{eqnarray}

\noindent which can be written in the more compact form

\begin{eqnarray}
\pdv{t} \hat{\rho}(t) &=& \gamma (\dyad{g}{e}\hat{\rho}(t)\dyad{e}{g} - \frac{1}{2}\dyad{e}{e}\hat{\rho}(t) -\frac{1}{2}\hat{\rho}(t)\dyad{e}{e}) \nonumber \\
&=& \frac{\gamma}{2} (2 \hat{\sigma} \hat{\rho}(t) \hat{\sigma}^{\dagger} - \hat{\sigma}^{\dagger}\hat{\sigma}\hat{\rho}(t) -\hat{\rho}(t)\hat{\sigma}^{\dagger}\hat{\sigma}) , \label{eq:optbloch_compact}
\end{eqnarray}

\noindent we can identify the collapse operator for spontaneous emission $\hat{L}_{spont}$ \cite{harocheraimond_spontemcollapse} by comparing eqn. \ref{eq:optbloch_compact} to the general form of the dissipator in the Lindblad equation:

\begin{eqnarray}
\hat{L}_{spont} = \sqrt{\gamma} \hat{\sigma}. \label{eq:spontem_collapse}
\end{eqnarray}

\noindent Note that the TLS lowering operator $\hat{\sigma} = \dyad{g}{e} = \frac{1}{2} (\hat{\sigma}_x + i \hat{\sigma}_y)$ can also be expressed in terms of the Pauli operators $\hat{\sigma}_x$ and $\hat{\sigma}_y$. Spontaneous emission is explicitly included into the formalism by inserting the collapse operator \ref{eq:spontem_collapse} into the Lindblad equation.

\section{Spectral and Laser Detuning in Rotating Frames} \label{sec:specdet}

\noindent Despite the finite linewidth of the transition, the center frequency $\omega_0 =\frac{E_e-E_g}{\hbar}$ is well defined through the energy eigenvalues of the ground and excited state. Two systems are considered spectrally detuned if their center frequencies $\omega_{01}$ and $\omega_{02}$ do not coincide. Photons emitted by dissimilar sources are thus generally subject to a spectral detuning $\Delta\omega = \omega_{02}-\omega_{01}$. Since in real experiments lasers may be slightly detuned from the transitions, we also account for laser detunings $\Delta_i = \omega_{Li}-\omega_{0i}$ in each system $i$ (with $\omega_{Li}$ the frequency of the excitation laser of system $i$). 

To include spectral and laser detuning we can explicitly incorporate the frequency of the emitter and the driving field into the Hamiltonian, which directly enters in the Lindblad equation \ref{eq:Lindblad_eq}. In a semiclassical picture and after performing the rotating wave approximation a Hamiltonian describing a driven TLS takes the form \cite{steck_TLS}:

\begin{eqnarray}
	\hat{H_i} = \hbar \omega_{0i} \hat{\sigma_i}^{\dagger} \hat{\sigma_i} + \frac{\hbar \Omega(t)}{2} \cdot (\hat{\sigma_i} e^{i \omega_{Li} t} + \hat{\sigma_i}^{\dagger} e^{-i \omega_{Li} t}) . \label{eq:TLS_hamiltonian_lab}
\end{eqnarray}

\noindent Eqn. \ref{eq:TLS_hamiltonian_lab} is defined in the laboratory frame. We can generally choose arbitrary reference frames for both system Hamiltonians. However, if we choose different coordinates to describe the dynamics of system 1 and 2, we have to transform the correlators \ref{eq:g2HOM_pulsed1}-\ref{eq:g2HOM_pulsed3} to a joint reference frame when merging them according to equation \ref{eq:g2HOM_pulsed}.

A common way to drastically reduce computational effort is by expressing the dynamics of each emitter in a respective rotating frame rotating at the laser frequency $\omega_{Li}$. In this way, all oscillatory time-dependence of the Hamiltonian in eqn. \ref{eq:TLS_hamiltonian_lab} is absorbed into the states and the Hamiltonian remains only time-dependent through the electric field amplitude. From the general rule for unitary transformations \cite{steck_TLS}

\begin{eqnarray}
&&\hat{\tilde{H}} = \hat{U}\hat{H}\hat{U}^{\dagger} + i \hbar (\pdv{t} \hat{U}) \hat{U}^{\dagger}  \label{eq:untrafH} \\
&&\ket*{\tilde{\psi}} = \hat{U} \ket*{\psi}, \label{eq:untrafPSI}
\end{eqnarray}

\noindent which for transforming the laboratory frame Hamiltonian $\hat{H}$ into a rotating frame Hamiltonian $\hat{\tilde{H}}$ requires the operator 

\begin{equation}
\hat{U} = e^{i \omega_{rot} t \dyad{e}{e}}, \label{eq: RF_unitary}
\end{equation}

\noindent with $\omega_{rot} \equiv \omega_{Li}$, we can infer the rule for transforming field operators into the rotating frame:

\begin{eqnarray}
\hat{\sigma}_i(t) &\rightarrow&  \hat{\sigma}_i(t)e^{-i\omega_{Li} t}     \label{eq:ladderop_transf1} \\
\hat{\sigma}_i^{\dagger}(t) &\rightarrow&  \hat{\sigma}^{\dagger}_i(t) e^{i\omega_{Li} t}    \label{eq:ladderop_transf2} \\
\mbox{with } \hat{H} &\rightarrow&  \hat{H} - \omega_{Li} \hat{\sigma}^{\dagger}_i\hat{\sigma}_i .
\end{eqnarray}

\noindent Using transformations \ref{eq:ladderop_transf1} and \ref{eq:ladderop_transf2} we can infer an expression for $g^{(2)}_{HOM}(0)$ that allows plugging in all correlators evaluated in their respective rotating frame. In this way, spectral and laser detuning appear in a $\tau$-dependent phase factor multiplied to the first-order correlation functions. Since similar phase factors cancel with their complex conjugate in the other terms, we only have to modify one term in equation \ref{eq:g2HOM_pulsed} to include spectral and laser detuning while working in rotating frames:

\begin{eqnarray}
\label{eq:g2HOM_pulsed_2rot}
&& 2 \Re{G^{(1)}_{11}(t,\tau)^* \cdot G^{(1)}_{22}(t,\tau)} \rightarrow \\
&& 2 \Re{G^{(1)}_{11}(t,\tau)_{RF_1}^* \cdot G^{(1)}_{22}(t,\tau)_{RF_2} \cdot e^{-i (\Delta \omega_0 +\Delta_2-\Delta_1) \tau}} . \nonumber 
\end{eqnarray}	

\noindent The index RF$_i$ indicates that the correlators are evaluated in respective reference frames rotating at $\omega_{Li}$. In the case of resonant excitation the phase factor reduces to $e^{-i \Delta \omega_0 \tau}$, and for emitters that differ only in emission frequency we get a real-valued factor of the form $\cos{(\Delta\omega_0 \tau)}$ \cite{citationkai}. 

Simulations of $g^{(2)}_{HOM}(0)$ for laser detunings up to 100 $\gamma$ in one or both emitters have shown that these detunings have a marginal influence on photon indistinguishability. Thus, in the following we set $\Delta \omega_2 =  \Delta \omega_1 \equiv 0$. 

\section{Temporal Delay} \label{sec:tempdel_app}

For four of the terms in eqn. \ref{eq:g2HOM_pulsed}, it can be shown by suitable substitutions of the time variables that they are not affected by a time delay between the photons:

\begin{widetext}
\begin{equation}
\begin{aligned}
G^{(2)}_{11}(t, \tau) &\xrightarrow{\delta\tau} G^{(2)}_{11}(t, \tau)  \\
G^{(2)}_{22}(t, \tau)  &\xrightarrow{\delta\tau} G^{(2)}_{22}(t-\delta\tau, \tau) \xrightarrow{t \rightarrow t- \delta \tau } G^{(2)}_{22}(t, \tau) \\
N_1(t)\cdot N_2(t+\tau) &\xrightarrow{\delta\tau} N_1(t)\cdot N_2(t-\delta\tau+\tau) \xrightarrow{\tau \rightarrow \tau- \delta \tau } N_1(t)\cdot N_2(t+\tau) \\
N_2(t)\cdot N_1(t+\tau) &\xrightarrow{\delta\tau} N_2(t-\delta\tau)\cdot N_1(t+\tau) \xrightarrow[]{t \rightarrow t- \delta \tau,~ \tau \rightarrow \tau+ \delta \tau} N_2(t)\cdot N_1(t+\tau) . \label{eq:timesubst}
\end{aligned}
\end{equation}
\end{widetext}

\section{Pure Dephasing Collapse Operator} \label{sec:puredeph_app}

\noindent Pure dephasing is accounted for in our formalism by including the corresponding collapse operator in eqn. \ref{eq:Lindblad_eq}. In the density matrix picture, pure dephasing corresponds to an approximately exponential decay of the coherences. This can be understood as a result of a statistical average of randomly z-rotated states on the Bloch sphere, leading to a mixed rather than a pure quantum state \cite{openquantumsystems_MIT}:
 
\begin{eqnarray}
&&\hat{\rho}_S(t)  \approx \mqty(\rho_{00} &  e^{-\gamma_{deph}t} \rho_{01} \\  e^{-\gamma_{deph}t} \rho_{10} & \rho_{11})\\
&&= \frac{1+e^{-\gamma_{deph}t}}{2}\cdot \hat{\rho}_S(0) + \frac{1-e^{-\gamma_{deph}t}}{2}\cdot \hat{\sigma}_z \hat{\rho}_S(0) \hat{\sigma}_z . \nonumber
\end{eqnarray}

\noindent By considering small time intervals dt, such that $e^{-\gamma_{deph}dt} \approx 1-\gamma dt$, we derive a differential equation for $\hat{\rho}(t)$:

\begin{eqnarray}
&&\lim_{dt\to0} \frac{\hat{\rho}_S(dt) - \hat{\rho}_S(0)}{dt} = \pdv{t} \hat{\rho}_S(t) \nonumber \\
&&= \frac{\gamma_{deph}}{2} (-\hat{\rho}_S(t) + \hat{\sigma}_z \hat{\rho}_S(t) \hat{\sigma_z}) . \label{eq:puredephasing_DGL}
\end{eqnarray}

\noindent Noting that $\hat{\sigma}_z = \hat{\sigma}^{\dagger}_z$ and $\hat{\sigma}_z \hat{\sigma}_z = \mathbb{1}$, eqn. \ref{eq:puredephasing_DGL} can be rewritten as

\begin{eqnarray}
\frac{\partial \hat{\rho}_S(t)}{\partial t} = \frac{\gamma_{deph}}{2} (\hat{\sigma}_z \hat{\rho}_S(t)\hat{\sigma}^{\dagger}_z -\frac{1}{2} \hat{\sigma}^{\dagger}_z\hat{\sigma}_z - \frac{1}{2} \hat{\rho}_S(t)\hat{\sigma}^{\dagger}_z\hat{\sigma}_z) . \label{eq:dephasing_dissipator}
\end{eqnarray}

\noindent Comparing eqn. \ref{eq:dephasing_dissipator} to the dissipator in the Lindblad equation (eqn. \ref{eq:Lindblad_eq}), we identify the collapse operator for pure dephasing to be

\begin{eqnarray}
\hat{L}_{deph} &=& \sqrt{\frac{\gamma_{deph}}{2}} \hat{\sigma}_z . \label{eq:dephasing_collapseop}
\end{eqnarray}

\section{Spectral Wandering Integral} \label{sec:specwan_app}

\noindent To get the distribution of spectral detunings we insert eqn. \ref{eq:wandering0} into eqn.\ref{eq:transfpdf}:

\begin{eqnarray}
p(\Delta \omega) &=& \frac{1}{2 \pi \sigma_1 \sigma_2} \int_{-\infty}^{\infty} \int_{-\infty}^{\infty} d\omega_1 d\omega_2 \delta(\omega_2 -( \Delta \omega + \omega_1)) \nonumber \\ && \cdot ~ e^{-\frac{1}{2 \sigma_1^2} (\omega_1-\omega_{01})^2} e^{-\frac{1}{2 \sigma_2^2}  (\omega_2-\omega_{02})^2}  \\
&=& \frac{1}{2 \pi \sigma_1 \sigma_2} \int_{-\infty}^{\infty} d\omega_1 e^{-\frac{1}{2 \sigma_1^2} (\omega_1-\omega_{01})^2}  e^{-\frac{1}{2\sigma_2^2}(\omega_1 + \Delta \omega - \omega_{02})^2} \\
&=&  \frac{1}{2 \pi \sigma_1 \sigma_2} e^{-\frac{1}{2}\frac{(\Delta \omega - \Delta \omega_0)^2}{\sigma_2^2}}  \int_{-\infty}^{\infty} dz e^{-\frac{1}{2}(\frac{1}{\sigma_1^2}+\frac{1}{\sigma_2^2}) z^2 - \frac{\Delta \omega - \Delta \omega_0}{\sigma_2^2}z} . \label{eq:integralwandering}
\end{eqnarray}	

\noindent In the last step we made the substitution $z=\omega_1 - \omega_{01}$ and defined $\Delta \omega_0 = \omega_{02}- \omega_{01}$. The integral in equation \ref{eq:integralwandering} can be solved analytically.

\section{Normalization} \label{sec:norm_app}

\noindent Depending on the experimental situation, an appropriately chosen normalization of $g^{(2)}_{HOM}(0)$ may be necessary to interpret the results. In its most general form in eqn. \ref{eq:ghomttau}, the HOM cross-correlation function corresponds to the joint probability density of having a photon at detector 1 at time $t$ and a second photon at detector 2 at time $t+\tau$. In its pulse-wise integrated form in eqn. \ref{eq:g2HOM_pulsed}, it gives the probability of having a photon at detector 1 and a photon at detector 2 at any time after exciting both emitters with a single pulse each. If only this coincidence probability is of interest, no normalization is required. However, in order to use $G^{(2)}_{HOM}(0)$ as a universal measure for photon indistinguishability, we need to consider the average number of photons arriving from each emitter. For ideal single photon emission from both systems, this mean number is one, making normalization redundant. In the general case of finite pulse widths leading to re-excitation, pulse areas deviating from $\pi$, finite laser detunings, or the presence of pure dephasing, the situation is different. If, for example, the photon emission probability is far smaller than one, most of the time no or at most one photon impinges on the beamsplitter. This leads to a low coincidence probability, even if the photons are fully distinguishable. Normalizing with the number of expected photons is required to restore the interpretation as a measure of indistinguishability. 

An elementary condition for a normalization term is that it yields $1$ in the case of unit photon emission probabilities in both systems. Moreover, it must become smaller if on average less than one photon is emitted in either system. The most natural way to achieve this is by choosing the term in an analogous way as for the degree of second order coherence $g^{(2)}(0)$ \cite{g1g2_scully}. Note that in the HOM case, we express the correlated fields $\hat{a}$ and $\hat{b}$ in terms of the input fields at the beamsplitter $\hat{a'}$ and $\hat{b'}$. Following this approach we obtain an intensity normalization term $\mathcal{N}_1$ which reads \cite{simulationpaper}:

\begin{eqnarray}
\mathcal{N}_1 &=& \expval*{\hat{a}^{\dagger}\hat{a}}\expval*{\hat{b}^{\dagger}\hat{b}}  \nonumber \\
&=& \frac{1}{2} \expval*{(\hat{a}'^{\dagger}-\hat{b}'^{\dagger})(\hat{a}'-\hat{b}')} \frac{1}{2} \expval*{(\hat{a}'^{\dagger}+\hat{b}'^{\dagger})(\hat{a}'+\hat{b}')} \nonumber \\ 
&=& \frac{1}{4} (\expval*{\hat{n}_{a'}}+\expval*{\hat{n}_{b'}})^2 . \label{eq:intnorm1}
\end{eqnarray}

\noindent A similar way to normalize is by using the mean intensity \cite{simulationpaper}:

\begin{eqnarray}
\mathcal{N}_2 = \frac{1}{2} (\expval*{\hat{n}_{a'}}^2+\expval*{\hat{n}_{b'}}^2) . \label{eq:intnorm2}
\end{eqnarray}

\noindent The difference between these approaches is that compared to eqn. \ref{eq:intnorm1}, a mixing term $\frac{1}{2} \expval*{\hat{n}_{a'}} \expval*{\hat{n}_{b'}}$ is missing in eqn. \ref{eq:intnorm2}. Both these approaches work well when the mean photon number is reduced in both systems simultaneously. However, both normalizations break down when one system emits with unit probability, while for the other system $\expval*{\hat{n}} \ll 1$. This is due to the fact that in this case $\mathcal{N}_1$ and $\mathcal{N}_2$ are lower bounded by 1/4 and 1/2, respectively. Therefore, at some point they fail to compensate for small coincidence probabilities due to the absence of photons. To overcome these limitations, we choose cross-polarization normalization as described in the main text.

\section{Full Equation of the Pulse-Wise Degree of HOM Coherence} \label{sec:fulleq}

\noindent Including all mechanisms mentioned in this work to the pulse-wise degree of HOM coherence, we obtain:

\begin{widetext}
\begin{eqnarray}
G^{(2)}_{HOM}(0) = \frac{1}{4}( 
\int_0^{\infty} \int_{-\infty}^{\infty} dt d\tau (\underbrace{G^{(2)}_{11}(t, \tau) + G^{(2)}_{22}(t, \tau)}_{multi-photon~emission}) + \int_0^{\infty} \int_{-\infty}^{\infty} dt d\tau (\underbrace{N_1(t)\cdot N_2(t+\tau) + N_2(t)\cdot N_1(t+\tau)}_{intensity}) \nonumber \\- \underbrace{\cos^2{(\phi)}}_{polarization} \int_0^{\infty} \int_{-\infty}^{\infty} dt d\tau~ 2 \Re{G^{(1)}_{11}(t,\tau)_{RF_1}^* \cdot G^{(1)}_{22}(t-\underbrace{\delta\tau}_{temporal~delay},\tau)_{RF_2} \cdot \underbrace{e^{-i (\Delta \omega+\Delta_2-\Delta_1) \tau}}_{spectral~and~laser~detuning}}).  \nonumber \label{eq:finaleq} \\ \nonumber
\end{eqnarray}	
\end{widetext}

\noindent Several effects enter implicitly through the time evolution of the Heisenberg operators and have to be incorporated to the Lindblad equation:

\begin{itemize}
    \item Spontaneous emission $\rightarrow$ 
    collapse operator $\sqrt{\gamma} \hat{\sigma}$
    \item Pure dephasing $\rightarrow$ collapse operator $\sqrt{\gamma_{deph}/2} \hat{\sigma_z}$
\end{itemize}

Other mechanisms have to be added supplementary:
\begin{itemize}
    \item Spectral wandering $\rightarrow$ mean value $\int_{-\infty}^{\infty} d\Delta \omega~ p(\Delta \omega) G^{(2)}_{HOM}(0, \Delta \omega)$
     \item Normalization $\rightarrow$ $g^{(2)}_{HOM}(0) \equiv G^{(2)}_{HOM}(0,\phi=0)/G^{(2)}_{HOM}(0,\phi=\pi)$
\end{itemize}

\noindent System type as well as parameters of the excitation mechanism enter into the Lindblad equation via the Hamiltonian. For time-resolved considerations, only integration over $\tau$ has to be omitted.


\bibliographystyle{apsrev4-2}
\bibliography{bibliography}

\end{document}